%% file: splitdoor_aoas_main.tex
\documentclass[arxiv,preprint]{imsart}

\RequirePackage[OT1]{fontenc}
\RequirePackage{amsthm,amsmath}
\RequirePackage{natbib}
\setcitestyle{square}
\RequirePackage[colorlinks,citecolor=blue,urlcolor=blue]{hyperref}

\RequirePackage{graphicx}
\usepackage{thmtools}
\usepackage{booktabs}
\usepackage{subfig}
\usepackage{centernot}
\usepackage{comment}
\usepackage[usenames]{xcolor}

\arxiv{arXiv:1611.09414}

\graphicspath{{figures/}}

\newcommand{\indep}{\mathrel{\text{\scalebox{1.07}{$\perp\mkern-10mu\perp$}}}}
\declaretheoremstyle[notefont=\bfseries,notebraces={}{},%
    headpunct={}]{mystyle}
\declaretheorem[style=mystyle,numbered=no,name=Assumption]{assumption-hand}

\startlocaldefs
\numberwithin{equation}{section}
\theoremstyle{plain}
\newtheorem{theorem}{Theorem}[section]
\newtheorem{lemma}{Lemma}
\newtheorem{assumption}{Assumption}
\endlocaldefs
\newenvironment{proof-intuit}{\paragraph{Proof (Argument)}}{\hfill}

\begin{document}

\begin{frontmatter}
\title{Split-door criterion: Identification of causal effects through auxiliary outcomes}
\runtitle{Split-door criterion for causal identification}
\thankstext{T1}{We would like to thank Dean Eckles, Praneeth Netrapalli, Joshua Angrist, T. Tony Ke, and anonymous reviewers for their valuable feedback on this work.}

\begin{aug}
    \author{\fnms{Amit} \snm{Sharma}\ead[label=e1]{amshar@microsoft.com}},
    \author{\fnms{Jake} M. \snm{Hofman}\ead[label=e2]{jmh@microsoft.com}}
    \and
    \author{\fnms{Duncan} J. \snm{Watts}
\ead[label=e3]{duncan@microsoft.com}
}

\runauthor{Sharma et al.}

\affiliation{Microsoft Research}

\address{9 Lavelle Road\\Bangalore, India 560008\\
\printead{e1}}
\address{641 Ave. of the Americas\\
New York, NY USA 10011 \\
\printead{e2}\\
\phantom{E-mail:\ }\printead*{e3}
}

\end{aug}

\begin{abstract}
We present a method for estimating causal effects in time series data when fine-grained information about the outcome of interest is available.
Specifically, we examine what we call the {\it split-door} setting, where the outcome variable can be split into two parts: one that is potentially affected by the cause being studied and another that is independent of it, with both parts sharing the same (unobserved) confounders.
We show that under these conditions, the problem of identification reduces to that of testing for independence among observed variables, and present a method that uses this approach to automatically find subsets of the data that are causally identified.
We demonstrate the method by estimating the causal impact of Amazon's recommender system on traffic to product pages, finding thousands of examples within the dataset that satisfy the split-door criterion.
Unlike past studies based on natural experiments that were limited to a single product category, our method applies to a large and representative sample of products viewed on the site.
In line with previous work, we find that the widely-used click-through rate (CTR) metric overestimates the causal impact of recommender systems; depending on the product category, we estimate that 50-80\% of the traffic attributed to recommender systems would have happened even without any recommendations.
We conclude with guidelines for using the split-door criterion as well as a discussion of other contexts where the method can be applied. 
\end{abstract}



\end{frontmatter}

\section{Introduction}
\label{sec:intro}

\begin{figure}[tb]
	\centering
    \subfloat[Canonical causal inference problem]{\includegraphics[scale=0.45]{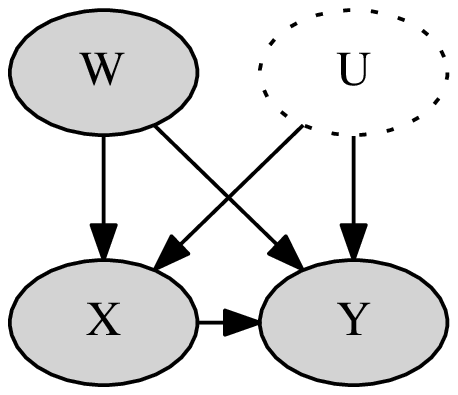}\label{fig:canonical-xy}}%
    \qquad
    \subfloat[Estimation with \textit{back-door} criterion]{\includegraphics[scale=0.45]{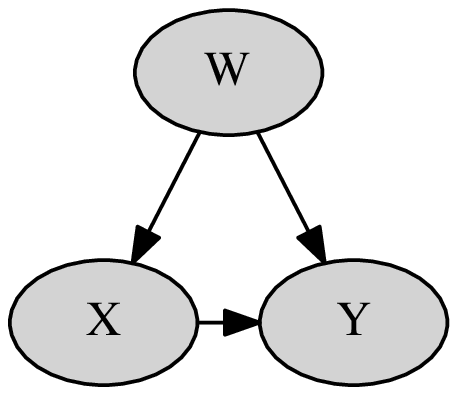}\label{fig:backdoor-xy}}%
	\qquad
    \subfloat[Estimation with Z as an \mbox{\textit{instrumental variable}}]{\includegraphics[scale=0.45]{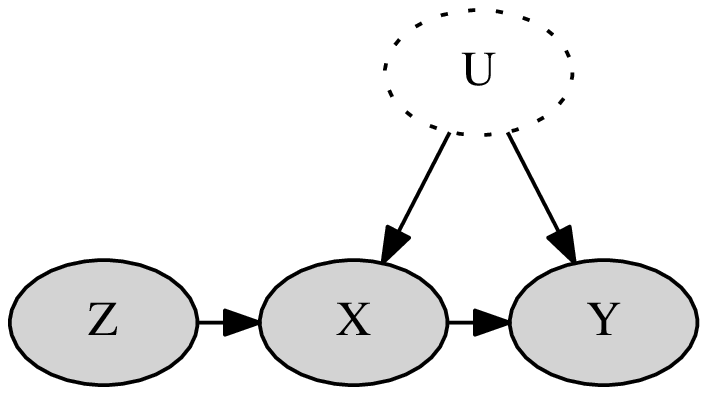}\label{fig:iv-xy}}%
	\qquad
	\caption{\textit{Left:} Graphical model for the canonical problem in causal inference. We wish to estimate the effect of $X$ on $Y$. $W$ represents observed common causes of $X$ and $Y$; $U$ represents other unobserved (and unknown) common causes that confound observational estimates. \textit{Middle}: The causal model under the \textit{selection on observables} assumption, where there are no known unobserved confounds $U$. \textit{Right:} The canonical causal model for an instrumental variable $Z$ that systematically shifts the distribution of the cause $X$ independently of confounds $U$.}
\end{figure}

The recent growth of digital platforms has generated an avalanche of highly granular and often longitudinal data regarding individual and collective behavior in a variety of domains of interest to researchers, including in e-commerce, healthcare, and social media consumption. Because the vast majority of this data is generated in non-experimental settings, researchers typically must deal with the possibility that any causal effects of interest are complicated by a number of potential confounds.
For example, even effects as conceptually simple as the causal impact of recommendations on customer purchases are likely confounded by selection effects~\citep{lewis2011}, correlated demand~\citep{sharma2015}, or other shared causes of both exposure and purchase.
Figure~\ref{fig:canonical-xy} shows this canonical class of causal inference problems in the form of a causal graphical model~\citep{pearl2009causality}, where X is the cause and Y is its effect. Together $U$ and $W$ refer to all of the common causes of X and Y that may  confound estimation of the causal effect, where critically some of these confounders (labeled $W$) may be observed, while others ($U$) are unobserved or even unknown. Ideally one would answer such questions by running randomized experiments on these platforms, but in practice such tests are possible only for the owners of the platform in question, and even then are often beset with implementation difficulties or ethical concerns~\citep{fiske2014}.
As a result researchers are left with two main strategies for making causal estimates from large-scale observational data, each with its own assumptions and limitations: either conditioning on observables or exploiting natural experiments.

\subsection{Background: Back-door criterion and natural experiments} 
The first and by far the more common approach is to assume that the effect of unobserved confounders ($U$) is negligible after conditioning on the observed variables ($W$). Under such a \textit{selection on observables} assumption \citep{imbens2015causal}, one conditions on $W$ to estimate the effect of $X$ on $Y$ when these confounders are held constant.
In the language of graphical models, this strategy is referred to as the \textit{back-door criterion} \citep{pearl2009causality} on the grounds that the ``back-door pathway'' from X to Y (via W) is blocked by conditioning on W (see Figure~\ref{fig:backdoor-xy})
and can be implemented by a variety of methods, including regression, stratification, 
 and matching~\citep{rubin2006matched,stuart2010matching}. Unfortunately for most practical problems it is difficult to establish that all of the important confounders have been observed.
 For example, consider the problem of estimating the causal impact of a recommender system on traffic to e-commerce websites such as Amazon.com, where $X$ corresponds to the number of visits to a product's webpage, and $Y$ the visits to a recommended product shown on that webpage.  One could compute the observed click-through rate after conditioning on all available user and product attributes (e.g., user demographics, product categories and popularities, etc.), assuming that these features constitute a proxy for latent demand.
 Unfortunately, there are also many potentially unobserved confounders (e.g., advertising, media coverage, seasonality, etc.) that impact both a product and its recommendations, which if excluded would render the back-door criterion invalid.

Motivated by the limitations of the back-door strategy, a second main approach is to identify an external event that affects the treatment $X$ in a way that is arguably random with respect to potential confounds.
The hope is that such variation, known as a \textit{natural experiment}~\citep{dunning2012natural}, can serve as a substitute for an actual randomized experiment.
Continuing with the problem of estimating the causal impact of recommendations, one might look for a natural experiment in which some products experience large and sudden changes in traffic, for instance when a book is featured on Oprah's book club~\citep{carmi2012}.
Assuming that the increase in traffic for the book is independent of demand for its recommendations, one can estimate the causal effect of the recommender by measuring the change in sales to the recommended products before and after the book was featured, 
arguing that these sales would not have happened in the absence of the recommender.
Such events provide 
\textit{instrumental variables} that identify the effect of interest by shifting the distribution of the cause $X$ independently of unobserved confounds $U$~\citep{angrist1996}.
Figure~\ref{fig:iv-xy} depicts this in a graphical model, where the additional observed variable $Z$ denotes the instrumental variable.

These two main approaches trade off critical goals of identification and generalization in causal inference.
The estimate for back-door conditioning is typically derived using all available data, but provides no identification guarantees in the presence of unobserved confounders.
Instrumental variables, in contrast, provide identification guarantees even in the presence of unobserved confounders, but 
these guarantees apply only for local subsets of the available data---the relatively rare instances for which a valid instrument that exogenously varies the cause $X$ is known (e.g., lotteries \citep{angrist1996}, variation in weather \citep{ phan2015}, or  sudden, large events \citep{rosenzweig2000,dunning2012natural}).

\begin{figure}[tb]
\centering
    \subfloat[General split-door model: Outcome $Y$ is split into $Y_R$ and $Y_D$]{
    	\includegraphics[scale=0.55]{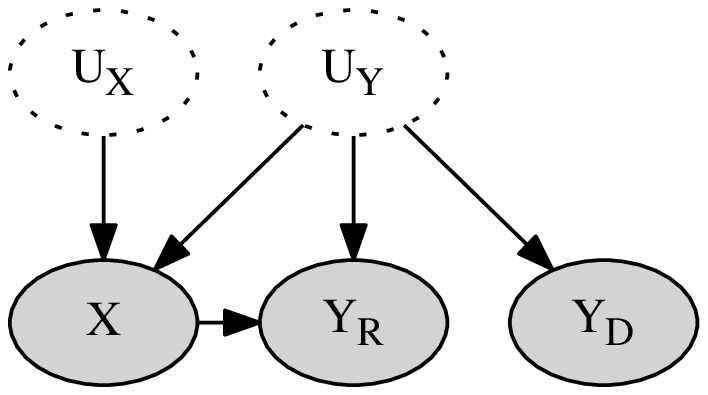}
    \label{fig:sdoor-model}}%
    \qquad
    \subfloat[Valid split-door model: Data subsets where $X$ is independent of $U_Y$]{
        \includegraphics[scale=0.55]{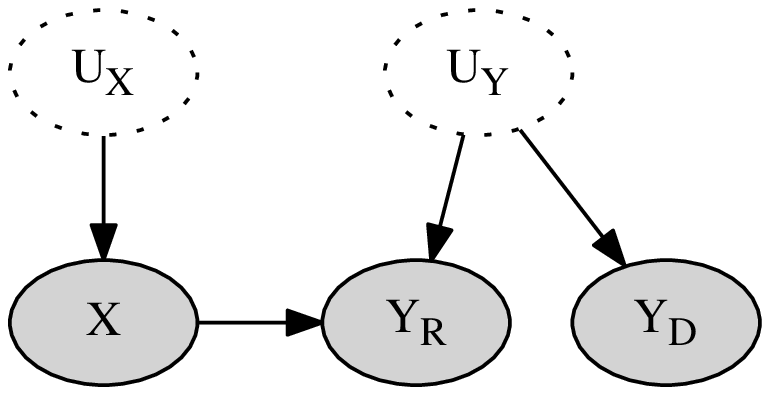}
    \label{fig:sdoor-model-natexp}}
    \caption{Panel (a) illustrates the canonical causal inference problem when outcome $Y$ can be split up into two components. For clarity,  unobserved confounders $U$ are broken into $U_Y$ that affects both $X$ and $Y$, and $U_X$ that affects only $X$. The split-door criterion finds subsets of the data where the cause $X$ is independent of $U_Y$ by testing independence of $X$ and $Y_D$, leading to the unconfounded causal model shown in Panel (b). }
\end{figure}

\subsection{The ``split-door'' criterion} 
In this paper we introduce a causal identification strategy that incorporates elements of both the back-door and natural experiment approaches, but that applies in a different setting.
Rather than conditioning on observable confounds $W$ or exploiting sources of independent variation in the cause $X$, we instead look to {\it auxiliary outcomes}~\citep{mealli2013} to identify subsets of the data that are causally identified.
Specifically, our strategy applies when the outcome variable $Y$ can be effectively ``split'' into two constituents: one that is caused by $X$ and another that is independent of it.
Figure~\ref{fig:sdoor-model} shows the corresponding causal graphical model, where $Y_R$ denotes the ``referred'' outcome of interest affected by $X$ and $Y_D$ indicates the ``direct'' constituent of $Y$ that does not directly depend on $X$.
Returning to the recommender system example, $Y_R$ corresponds to recommendation click-throughs on a product whereas $Y_D$ would be all other traffic to that product that comes through channels such as direct search or browsing.
Whenever such fine-grained data on $Y$ is available, we show that it is possible to reduce causal identification to an independence test between the cause $X$ and the auxiliary outcome $Y_D$.
Because this strategy depends on the availability of a split set of variables for $Y$, we call it  the \textit{split-door} criterion for causal identification, by analogy with the more familiar back-door criterion.

Although we make no assumptions about the functional form of relationships between variables, a crucial assumption underlying the split-door criterion is {\it connectedness}; i.e., that the auxiliary outcome $Y_D$ must be affected (possibly differently) by {\it all} causes that also affect $Y_R$.
As we discuss in more detail in Section~\ref{sec:amazon}, this assumption is plausible in scenarios such as online recommender systems, where recommended products are reachable through multiple channels (e.g., search or direct navigation) and it is unlikely that demand for a product manifests itself exclusively through only one of these channels. More generally, the connectedness assumption is expected to hold in scenarios where direct and referred outcomes incur similar cost, which makes it unlikely that something that causes the outcome does so 
only when referred through $X$, but never directly.

Under the above assumption, the split-door criterion seeks to identify subsets of the data where causal identification is possible.
In this sense, the method resembles a natural experiment, except that instead of looking for an instrument that creates variation in $X$, we look for variations in $X$ directly.
As in a natural experiment, however, it is important that any such variation in $X$ is independent of potential confounds.
For instance in the example above, it is important that a sudden burst of interest in a particular book is not correlated with changes in latent demand for its recommendations.
To verify this requirement, the split-door criterion relies on a statistical test to select for cases where there are no confounds (observed or otherwise) between $X$ and $Y_R$.
Specifically, we show that given a suitable auxiliary outcome $Y_D$, and a test to establish if $X$ and $Y_D$ are independent, the causal effect between $X$ and $Y_R$ can be identified.
Furthermore, since this test involves two observed quantities ($X$ and $Y_D$), we can systematically search for subsets of the data that satisfy the required condition, potentially discovering a large number of cases in which we can identify the causal effect of $X$ on $Y_R$.

We illustrate this method with a detailed example in which we estimate the causal impact of Amazon.com's recommendation system using historical web browsing data.
Under the above assumptions on the dependence between referred and direct visits to a product's webpage, we show how the criterion provides a principled mechanism for determining which subsets of the data to include in the analysis.
The split-door criterion identifies thousands of such instances in a nine-month period, comparable in magnitude to a manually tuned approach using the same data \citep{sharma2015}, and an order of magnitude more than traditional approaches \citep{carmi2012}.
Further, the products included in our analysis are representative of the overall product distribution over product categories on Amazon.com, thereby improving both the precision and generalizability of estimates.
Consistent with previous work \citep{sharma2015}, we find that observational estimates of recommendation click-through rates (CTRs) overstate the actual effect by anywhere from 50\% to 80\%, calling into question the validity of popular CTR metrics for assessing the impact of recommendation systems.
For applications to other online and offline scenarios, we provide an R package\footnote{URL: \url{http://www.github.com/amit-sharma/splitdoor-causal-criterion}} that implements the split-door criterion.

\subsection{Outline of paper}
The remainder of this paper proceeds as follows.
In Section~\ref{sec:the split-door} we start with a formal definition of the split-door criterion and give precise conditions under which the criterion holds.
For clarity we provide proofs for causal identification both in terms of the causal graphical model from Figure~\ref{fig:sdoor-model} and also in terms of structural equations. 
In Section~\ref{sec:splitdoor-algorithm} we propose a simple, scalable algorithm for identifying causal effects using the split-door criterion.
Then in Section~\ref{sec:splitdoor-compare}, we explain more formally how the split-door criterion differs from the instrumental variables and back-door methods mentioned above.
Section~\ref{sec:amazon} presents details about the Amazon.com data and an application of the split-door criterion to estimate the causal impact of its recommendation system. 
In Section~\ref{sec:discussion} we then discuss limitations of the split-door criterion as well as other settings in which the criterion applies, arguing that many existing datasets across a variety of domains have the structure that outcomes of interest can be decomposed into their ``direct'' and ``referred'' constituents.
We conclude with a prediction that as the size and granularity of available datasets, along with the number of variables in them, increase at an ever faster rate,  data-driven approaches to causal identification will become commonplace.

\input{splitdoor_details}

\input{amazonrecs_application}

\input{discussion}
\input{appendix}

\bigskip






\begin{supplement}
\sname{Supplement A}\label{suppA}
\stitle{Code for split-door criterion}
\slink[url]{http://www.github.com/amit-sharma/splitdoor-causal-criterion}
\sdescription{We provide an R package that implements the split-door criterion, along with code samples for applying the criterion to new applications.}
\end{supplement}

\bibliographystyle{imsart-nameyear}

\bibliography{causal_inference}

\end{document}

%% file: splitdoor_details.tex
\section{The Split-door Identification Criterion}
\label{sec:the split-door}
 
The split-door criterion can be used whenever observed data is generated from the model shown in Figure~\ref{fig:sdoor-model}. 
Here $X$ represents the cause of interest, $Y_R$ denotes the ``referred'' portion of the outcome affected by it, and $Y_D$ indicates the ``direct'' part of the outcome which does not directly depend on $X$.
We denote the overall outcome by $Y = Y_R + Y_D$.
We let $U_Y$ represent all unobserved causes of $Y$, some of which may also be common causes of $X$, hence the arrow from $U_Y$ to $X$.
Additional latent factors that affect only $X$ are captured by $U_X$.
Both $U_X$ and $U_Y$ can be a combination of many variables, some observed and some unobserved. (For full generality, the analysis presented here assumes that all confounds are unobserved.)
As noted earlier, the unobserved variables $U_Y$ create ``back-door pathways'' that confound the causal effect of $X$ on $Y$, resulting in biased estimates.
The central idea behind the split-door criterion is that we can use an independence test between the auxiliary outcome $Y_D$ and $X$ to systematically search for subsets of the data that are free of these confounds and do not contain back-door pathways between $X$ and $Y_R$.
In other words, we can conclude that such subsets of the data were generated from the unconfounded causal model shown in Figure~\ref{fig:sdoor-model-natexp}, and therefore the causal effect of $X$ on $Y$ can be estimated directly from these data.
Importantly, identification of the causal effect rests on the assumption that no part of $U_Y$ causes one part of $Y$ and not the other.

\subsection{The split-door criterion through a graphical model}
\label{sec:splitdoor-graphical}
Here we formalize the intuition above in the causal graphical model framework.
To identify the causal effect, we make the following two assumptions.
The first pertains to connectedness of the causal model.

\begin{assumption}[Connectedness] \label{as:connectedness}
\textit{Any unobserved confounder $U_Y$ that causes both $X$ and $Y_R$ also causes $Y_D$ and the causal effect of such $U_Y$ on $Y_D$  is non-zero.
}\end{assumption}

Note that Assumption 1 requires only that the causal effect of $U_Y$ on $Y_D$ be non-zero, without any requirements on the size of the effect(s) involved.
That said, it is a strong requirement in general, as it applies to all sub-components of $U_Y$ and thus involves assumptions about potentially high-dimensional, unobserved variables.
Whenever $Y_D$ and $Y_R$ are components of the same variable it is plausible that they share causes, but one still must establish that this condition holds to ensure causal identification.
It is instructive to compare this assumption to the strict independence assumptions involving unobserved confounders required by methods such as instrumental variables \citep{angrist1996}. 

The second assumption, which relates statistical and causal independence between observed variables, is standard for many methods of causal discovery from observational data.

\begin{assumption}[Independence]
\textit{If $X$ and $Y_D$ are statistically independent, then they are also causally independent in the graphical model of Figure~\ref{fig:sdoor-model}.
}\end{assumption}

Here \emph{causal} independence between two variables means that they share no common causes and no directed path in the causal graphical model leads from one to another. More formally, the two variables are \emph{``d-separated''} \citep{pearl2009causality} from each other. Thus, Assumption 2 is a variant of the Faithfulness or Stability assumptions in causal graphs with latent unobserved variables~\citep{spirtes2000causation,pearl2009causality}.
In the causal model shown in Figure~\ref{fig:sdoor-model}, for instance, this assumption rules out the possibility of an event where the observed variables $X$ and $Y_D$ are found to be statistically independent, but $U_Y$ still affects both of them and the observed independence in the data results from $U_Y$'s effect canceling out exactly over the path $X$-$U_Y$-$Y_D$. 
In other words, this assumption serves to rule out an (unlikely) event where incidental equality of parameters or certain data distributions render two variables statistically independent even though they are causally related.

Under Assumptions 1 and 2, 
 we can show that statistical independence of $X$ and $Y_D$ ensures that $X$ is not confounded by $U_Y$. First, we provide a result about the resulting causal graph structure when $X \indep Y_D$.

\begin{lemma}
\label{thm:splitdoor-lemma}
Let $X$, $Y_R$ and $Y_D$ be three observed variables corresponding to the causal model in Figure~\ref{fig:sdoor-model}, where $U_Y$ refers to unobserved causes of $Y_R$. If the \textit{connectedness} (1) and \textit{independence} (2) assumptions hold, then $X \indep Y_D$ implies that  the edge $U_Y \rightarrow X$ does not exist or that $U_Y$ is constant. 
\end{lemma}

\begin{proof-intuit}
The proof can be completed directly from Figure~\ref{fig:sdoor-model} and properties of a causal graphical model.

$X \indep Y_D$ implies that the causal effect of $U_Y$ on $Y_D$ and $X$ somehow cancels out on the path $X \leftarrow U_Y \rightarrow Y_D$. By \textit{Assumption 2}, this cancellation is not due to incidental equality of parameters or a particular data distribution, but rather a property of the causal graphical model. Therefore, this can only happen if \\ 
(i) $U_Y$ is constant (and thus \textit{blocks} the path), or \\
(ii) One of the edges exists trivially (does not have a causal effect). Using \textit{Assumption 1}, $U_Y$ has a non-zero effect on  $Y_D$. Then, the only alternative is that the $X \leftarrow U_Y$ edge does not exist, leading to the unconfounded causal model in Figure~\ref{fig:sdoor-model-natexp}.
\end{proof-intuit}
\begin{proof}
We provide a proof by contradiction using the principle of \textit{d-separation} \citep{pearl2009causality} in a causal graphical model.

Let us suppose $X \indep Y_D$, and that the $U_Y \rightarrow X$ edge exists and $U_Y$ is not constant.

Using the rules of \textit{d-separation} on the causal model in Figure~\ref{fig:sdoor-model}, the path $X$-$U_Y$-$Y_D$ corresponds to:
\begin{align}
&(X \indep Y_D|U_Y)_G \\
&(X \not \indep Y_D)_G
\end{align}
where the notation $(.)_G$ refers to \textit{d-separation} under a causal model $G$. In our case, $G$ corresponds to the causal model in Figure~\ref{fig:sdoor-model}.

However, using Assumption 2, statistical independence of $X$ and $Y_D$ implies causal independence, and thus, d-separation of $X$ and $Y_D$.
\begin{align}
(X\indep Y_D)_G
\end{align}

Equations 2.2 and 2.3 result in a contradiction. To resolve,  

(i) Either $U_Y$ is constant and thus 2.1 implies $(X \indep Y_D)_G$ holds, or 

(ii) The path $X$-$U_Y$-$Y_D$ does not exist. Using \textit{Assumption 1} of dependence of $Y_D$ on $U_Y$, the only possibility is that the $X \leftarrow U_Y$ edge does not exist.

\end{proof}

We now show that Lemma~\ref{thm:splitdoor-lemma} removes confounding due to $U_Y$ and that the observational estimate $P(Y_R|X=x)$ is also the causal estimate. 

\begin{theorem}[Split-door Criterion] \label{thm:splitdoor-criterion}
Under the assumptions of Lemma~\ref{thm:splitdoor-lemma}, the causal effect of $X$ on $Y_R$ is not confounded by $U_Y$ and is given by:
$$P(Y_R|do(X=x)) = P(Y_R|X=x)$$
where $do(X=x)$ refers to experimental manipulation of $X$ and $Y_R|X=x$ refers to the observed conditional distribution.
\end{theorem}
\begin{proof-intuit}
Lemma~\ref{thm:splitdoor-lemma} leads to two cases:\\
(i)  By the back-door criterion \citep{pearl2009causality}, if $U_Y$ is constant, then $X$ and $Y_R$ are unconfounded, because the only back-door path between $X$ and $Y_R$ contains $U_Y$ on it. \\
(ii) Similarly, if the $U_Y \rightarrow X$ edge does not exist, then $X$ and $Y_R$ are unconfounded because absence of the $U_Y \rightarrow X$ edge removes the back-door path between $X$ and $Y_R$. 

In both cases, unconfoundedness implies that the effect of $X$ on $Y_R$ can be estimated using the observational distribution.
\end{proof-intuit}
\begin{proof}
The proof follows from an application of the second rule of do-calculus \citep{pearl2009causality}.
\begin{equation}
P(\mathcal{Y}|do(\mathcal{Z}=z),\mathcal{W}) = P(\mathcal{Y}|\mathcal{Z}=z, \mathcal{W}) \text{\ \ \ \ \ if \ \ \ \ \ } (\mathcal{Y} \indep \mathcal{Z}|\mathcal{W})_{G_{\underline{\mathcal{Z}}}}
\label{eqn:rule-do}
\end{equation}
where $G_{\underline{Z}}$ refers to the underlying causal graphical model with all outgoing edges from $\mathcal{Z}$ removed.

Substituting $\mathcal{Y}=Y_R$, $\mathcal{Z}=X$,  $G_{\underline{X}}$ corresponds to the causal model from Figure~\ref{fig:sdoor-model} without the $X \rightarrow Y_R$ edge.  Using Lemma~\ref{thm:splitdoor-lemma}, two cases exist:

(i) $U_Y$ is constant \\ 
Let $\mathcal{W}=U_Y$. Under the modified causal model $G_{\underline{X}}$ without the $X \rightarrow Y_R$ edge, the path $X$-$U_Y$-$Y_R$ is the only path connecting $X$ and $Y_R$, which leads to the following \textit{d-separation} result:
\begin{align}
&(Y_R \indep X|U_Y)_{G_{\underline{X}}}
\end{align}

Combining Rule~\ref{eqn:rule-do} and the above \textit{d-separation} result, we obtain
\begin{equation*}
P(Y_R|do(X=x), U_Y)=P(Y_R|X=x, U_Y) = P(Y_R|X=x)
\end{equation*}
where the last equality holds because $U_Y$ is constant throughout.

(ii) The edge $U_Y \rightarrow X$ does not exist. \\
Let $\mathcal{W}=\emptyset$. Under the modified causal model $G_{\underline{X}}$ without the $X \rightarrow Y_R$ edge, $X$ and $Y_R$ are trivially \textit{d-separated} because no path connects them without the edge $U_Y \rightarrow X$.
\begin{equation}
(Y_R \indep X)_{G_{\underline{X}}}
\end{equation}

From Rule~\ref{eqn:rule-do} and the above \textit{d-separation} result, we obtain
\begin{equation*}
P(Y_R|do(X=x)) = P(Y_R|X=x)
\end{equation*}

\end{proof}

\subsection{The split-door criterion through structural equations}
\label{sec:splitdoor-struc-eqns}
Although we have already analyzed the split-door criterion in terms of the causal graphical model in Figure~\ref{fig:sdoor-model}, for expositional clarity we note that it is also possible to do the same using structural equations. Specifically, we can write three structural equations:
\begin{align}
    x  = g(u_x,u_y, \varepsilon _x) & &
    y_r  = f(x, u_y, \varepsilon _{yr}) & &
	y_d  = h(u_y, \varepsilon _{yd}),
\end{align}
where $\varepsilon _x$, $\varepsilon _{yr}$,  and $\varepsilon _{yd}$ are mutually independent, zero-mean random variables that capture modeling error and statistical variability. As in \textit{Assumption 1}, 
we assume that $U_Y$ affects both $Y_D$ and $Y_R$. 
In general, the causal effects among variables may not be linear; however, for the purpose of building intuition we rewrite the above equations in linear parametric form:
\begin{align} \label{eqn:linear-struc-eqns1}
       x  = \eta u_x + \gamma_1 u_y + \epsilon _x & &
       y_r  = \rho x + \gamma_2 u_y + \epsilon _{yr} & &
y_d  = \gamma_3 u_y + \epsilon _{yd}, 
\end{align}
where $\rho$ is the causal parameter of interest, and $\epsilon _x$, $\epsilon _{yr}$ $\epsilon _{yd}$ are independent errors in the regression equations.  The split-door criterion requires independence of $X$ and $Y_D$, which in turn implies that $\operatorname{Cov}(X, Y_D) = 0$:
\begin{equation*}
\begin{split}
    0 = \operatorname{Cov}(X, Y_D) & = \operatorname{E}[XY_D] - \operatorname{E}[X]\operatorname{E}[Y_D] \\
                                   & = \operatorname{E}[(\eta u_x+\gamma_1 u_y + \epsilon _x)(\gamma_3 u_y + \epsilon _{yd})] 
    - \operatorname{E}[\eta u_x+\gamma_1 u_y + \epsilon _x]\operatorname{E}[\gamma_3 u_y + \epsilon _{yd}] \\
    & = \gamma_1 \gamma_3 \operatorname{E}[U_Y.U_Y] - \gamma_1 \gamma_3 \operatorname{E}[U_Y]\operatorname{E}[U_Y]  \\
            & = \gamma_1 \gamma_3 \operatorname{Var}(U_Y)	
\end{split}
\end{equation*}
Assuming that $Y_D$ is affected by $U_Y$ (and therefore $\gamma_3$ is not $0$), the above can be zero only if $\gamma_1=0$, or if $U_Y$ is constant ($\operatorname{Var}[U_Y] = 0$). In both cases, $X$ becomes independent of $U_Y$ and the following regression can be used as an unbiased estimator for the  effect of $X$ on $Y_R$:
\begin{equation}
y_r  = \rho x + \epsilon' _{yr} \\
\end{equation}
where $\epsilon'_{yr}$ denotes an independent error.

\section{Applying the Split-door Criterion}
\label{sec:splitdoor-algorithm}
The results of the previous section motivate an algorithm for applying the split-door criterion to observational data.
Specifically, given an empirical test for independence between the cause $X$ and the auxiliary outcome $Y_D$, we can select instances in our data that pass this test and satisfy the split-door criterion.
In this section we develop such a test for time series data, resulting in a simple, scalable identification algorithm.

At a high level, the algorithm  works as follows.
First, divide the data into equally-spaced time periods $\tau$ such that each period has enough data points to reliably estimate the joint probability distribution $P(X, Y_D)$. Then, for each time period $\tau$,
\begin{enumerate}
\item Determine whether $X$ and $Y_D$ are independent using an empirical independence test.
\item If $X$ and $Y_D$ are determined to be independent, then the current time period $\tau$ corresponds  to a valid split-door \emph{instance}.  Use the observed conditional probability $P(Y_R|X=x)$ to estimate the causal effect in the time period $\tau$. Otherwise, exclude the current time period  from the analysis.
\item Average over all time periods where $X \indep Y_D$ to obtain the mean causal effect of $X$ on $Y_R$.
\end{enumerate}
Implementing the algorithm requires making suitable choices for an independence test and also its significance level, taking into account multiple comparisons. 
In the following sections, we discuss these choices in detail, as well as sensitivity of the method to violations in our assumptions.

\subsection{Choosing an independence test}
\label{sec:choose-indep-test}
Each $X$-$Y_D$ pair in \textit{Step 1} provides two vectors of length $\tau$ with observed values for $X$ and $Y_D$.
The key decision is whether these vectors are independent of each other.
In theory any empirical test that reliably establishes independence between $X$ and $Y_D$ is sufficient to identify instances where the split-door criterion applies.
For instance, assuming we have enough data, we could test for independence by comparing the empirical mutual information to zero \citep{steuer2002mutual,pethel2014exact}. 
In practice, however, because we consider subsets of the data over relatively small time periods $\tau$, there may be substantial limits to the statistical power we have in testing for independence.
For example, it is well known that in small sample sizes, testing for independence via mutual information estimation can be heavily biased \citep{paninski2003mutinfo}. 

Thus, when working with small time periods $\tau$ we recommend the use of exact independence tests and randomization inference~\citep{agresti1992survey,agresti2001exact,lydersen2007choice}.\footnote{
  When $X$ and $Y_D$ are discrete variables, methods such as Fisher's exact test are appropriate.
  If, however, $X$ and $Y_D$ are continuous---as is this case for the example we study in Section~\ref{sec:amazon}---we recommend the use of resampling-based randomization inference for establishing independence.
}
In general, this approach involves repeatedly sampling randomized versions of the empirical data to simulate the null hypothesis and then comparing a test statistic on the observed data to the same on the null distribution.
Specifically, for each $X$-$Y_D$ pair, we simulate the null hypothesis of independence between $X$ and $Y_D$ by replacing the observed $X$ vector with a randomly sampled vector from the overall empirical distribution of $X$ values.
From this simulated $X$-$Y_D$ instance, we compute a test statistic that captures statistical dependence, such as the distance correlation, which can detect both non-linear and linear dependence~\citep{szekely2007dcor,santos2014distance}.
We then repeat this procedure many times to obtain a null distribution for the test statistic of this $X$-$Y_D$ pair. 
Finally, we compute the probability $p$ of obtaining a test statistic as extreme as the observed statistic under the null distribution, and select instances in which the probability $p$ is above a pre-chosen significance level $\alpha$.

\subsection{Choosing a significance level}
\label{sec:siglevel}
In contrast to standard hypothesis testing where one is looking to reject the null hypothesis that two variables are independent and therefore thresholds on a small $p$-value, here we are looking for independent $X$-$Y_D$ pairs that are highly probable under the null and thus want a large $p$-value.
In other words, we are interested in a low Type II error (or false negatives), in contrast to standard null hypothesis testing, where the focus is on Type I errors (false positives) and hence significance levels are set low. Therefore, one way to choose a significance level would be to choose $\alpha$ as close as possible to 1 to minimize Type II errors when $X$ and $Y_D$ are dependent. 
At the same time, we need to ensure that the test yields adequate power for finding independent $X$-$Y_D$ pairs.
Unlike a conventional hypothesis test for dependent pairs, power for our test is $1-\alpha$, the probability that the test declares an $X$-$Y_D$ pair to be independent when it is actually independent.
As we increase $\alpha$, type II errors decrease, but power also decreases. 

Complicating matters, the combination of low power and a large number of hypothesis tests raises concerns about falsely accepting pairs that are actually dependent. As an extreme example, even when all $X$-$Y_D$ pairs in a given dataset are dependent, some of them will pass the independence test simply due to random chance. Therefore, a more principled approach to selecting $\alpha$ comes through estimating the expected fraction of erroneous split-door instances returned by the procedure, which we refer to as $\phi$.
As described in Appendix \ref{sec:fdr}, we apply techniques from the multiple comparisons literature~\citep{storey2002direct,liang2012fdr,farcomeni2008fdrreview} to estimate this fraction $\phi$ for any given significance level.

\subsection{Sensitivity to identifying assumptions}
\label{sec:sens-summary}
The above algorithm yields a causal estimate only if the identifying assumptions of \emph{connectedness} and \emph{independence} are satisfied. Independence is based on the standard faithfulness assumption in causal discovery \citep{spirtes2000causation}. Connectedness, on the other hand, requires justification based on domain knowledge. Even when the connectedness assumption seems plausible, we recommend a sensitivity analysis to assess the effects of potential violations to this assumption.

\begin{figure}[tb]
\centering
    \subfloat[General model: Unobserved variables split into $U_Y$ and $V_Y$]{
    	\includegraphics[scale=0.45]{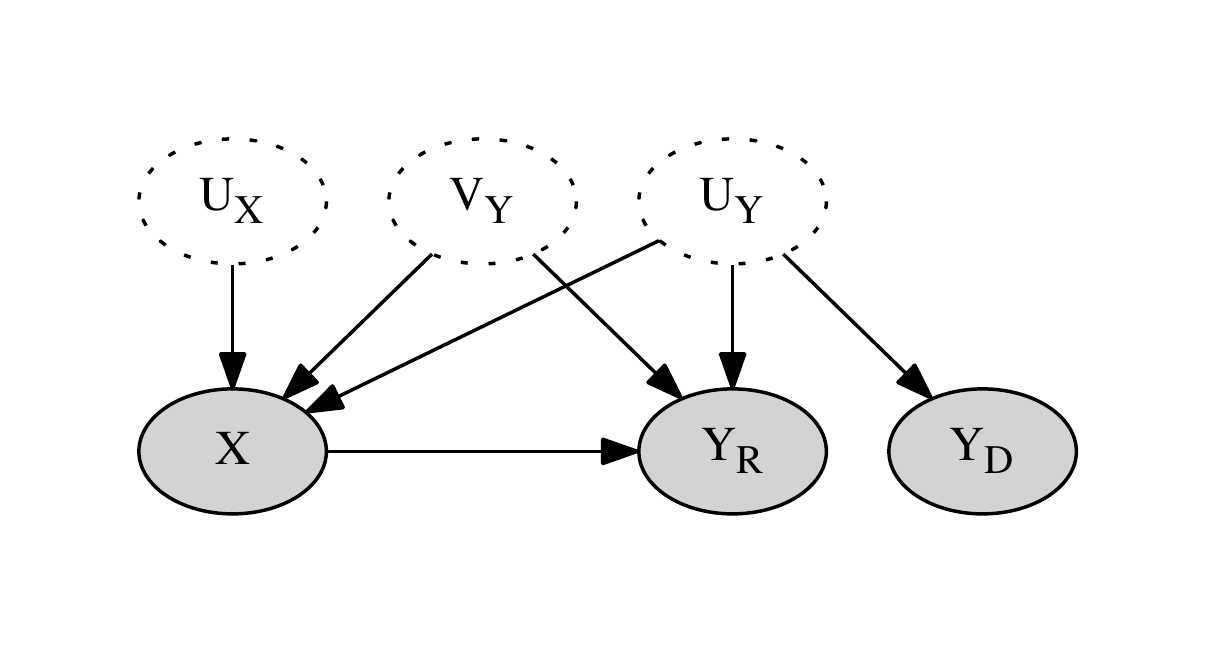}
    \label{fig:sens-connected1}}%
    \qquad
    \subfloat[Invalid split-door model: $X$ is independent of $U_Y$ but not $V_Y$]{
        \includegraphics[scale=0.45]{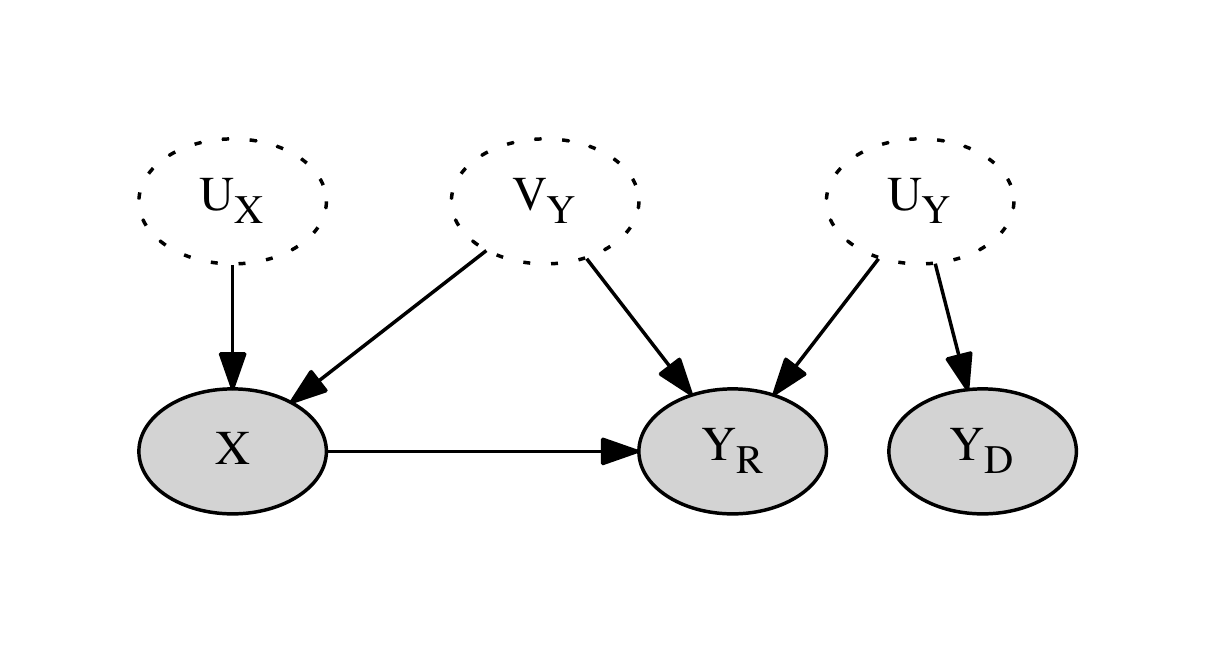}
    \label{fig:sens-connected2}}
    \caption{Violation of the connectedness assumption. Causes for $X$ and $Y_R$ consist of two components, $U_Y$ and $V_Y$, where $V_Y$ does not affect $Y_D$ and hence is undetectable by the split-door criterion. In the general causal model shown in Panel (a), $X\rightarrow Y_R$ is confounded by both $U_Y$ and $V_Y$. In the causal model corresponding to a split-door instance in Panel (b), $X\rightarrow Y_R$ is still confounded by the common cause $V_Y$. 
    }
\end{figure}

From Assumption \emph{1}, violation of connectedness implies that there exist some unobserved variables that affect $X$ and $Y_R$ but not $Y_D$.
Figure~\ref{fig:sens-connected1} shows this scenario, which is identical to the model in Figure~\ref{fig:sdoor-model} with the addition of an unobserved variable $V_Y$ that affects $X$ and $Y_R$, but not $Y_D$.
Applying the split-door criterion in this setting ensures that there is no effect of $U_Y$ on $X$, but does not alleviate possible confounds from $V_Y$, as shown in Figure~\ref{fig:sens-connected2}.
Note that this is analogous to the situation in back-door-based methods when one fails to condition on unobserved variables that affect both the treatment and outcome.
Correspondingly, sensitivity analyses designed for back-door-based methods \citep{harding2009sensitivity,rosenbaum2010design,vanderweele2011bias,carnegie2016sensitivity} can be readily adapted to analyzing split-door instances.
In addition, noting that split-door estimates represent averages over all discovered split-door instances, we introduce an additional sensitivity parameter $\kappa$ that denotes the fraction of instances for which connectedness is violated.
In Appendix~\ref{sec:sensitivity} we provide a derivation showing that sensitivity for the split-door estimate reduces to sensitivity for back-door methods and conduct this analysis for the application presented in Section~\ref{sec:amazon}.

\section{Connections to other methods}
\label{sec:splitdoor-compare}
The split-door criterion is an example of methods that use empirical independence tests to identify causal effects 
under certain assumptions \citep{jensen2008automatic,cattaneo2015randomization,sharma2015,grosse2016identification}. By searching for subsets of the data where desired independence holds, it also shares some properties with natural experiment methods such as instrumental variables and conditioning methods such as regression. We discuss these connections below; table~\ref{tab:compare-models} provides a summary for easy comparison.

\begin{table}[t]
     \centering
     \tiny{
     \begin{tabular}{ p{2.25cm} p{2cm} p{1.5cm}  p{1.9cm} p{2.25cm}  }
     \toprule
      Graphical model & Description & Untestable assumptions & Limitations & Recommendations example\\ 
    \cmidrule(r){1-1}\cmidrule(lr){2-4}\cmidrule(l){5-5}
     \begin{minipage}{.2\textwidth}
     \centering
      \includegraphics[height=1.25cm]{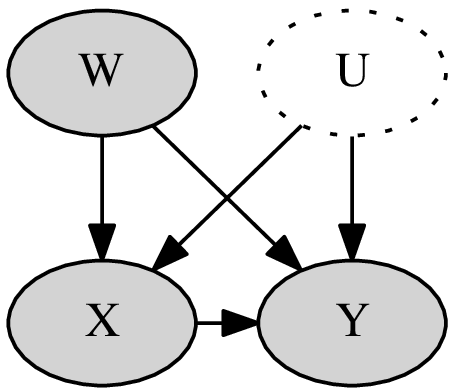}
  \end{minipage} \newline \textbf{(a)} \raggedright{Back-door criterion} &
      \vspace{-0.5cm}
       Condition on observed confounders $W$ to isolate the treatment effect. & \vspace{-0.5cm} $X \indep U$ or $Y \indep U$ & \vspace{-0.5cm} Unlikely that there are no unobserved confounders $U$. & \vspace{-0.5cm} Regress click-throughs on product attributes and direct visits to recommended product.\\ 
  \begin{minipage}{.2\textwidth}
  	\centering
      \includegraphics[height=1.25cm]{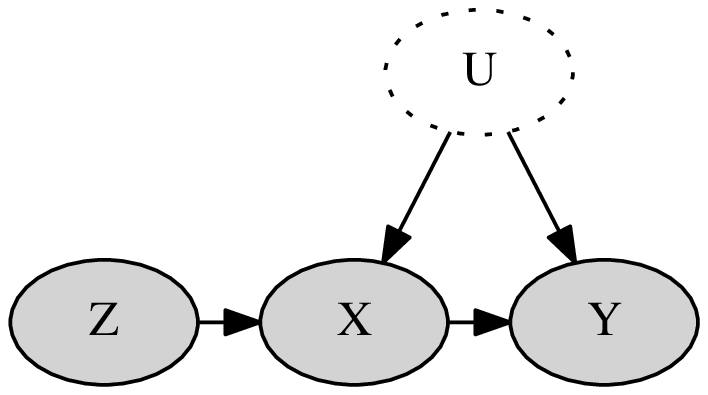}
  \end{minipage}  \newline \textbf{(b)} \raggedright{Instrumental variable} &
      \vspace{-0.5cm}  Analyze subset of data that has independent variation in the treatment.  & \vspace{-0.5cm} $Z \indep U$ and $Z \indep Y | X, U$ & \vspace{-0.5cm} Difficult to find a source of exogenous variation in the treatment. & \vspace{-0.5cm} Measure marginal click-throughs on products that experience large, sudden shocks in traffic. \\ 

  \begin{minipage}{.2\textwidth}
	  \centering
      \includegraphics[height=1.25cm]{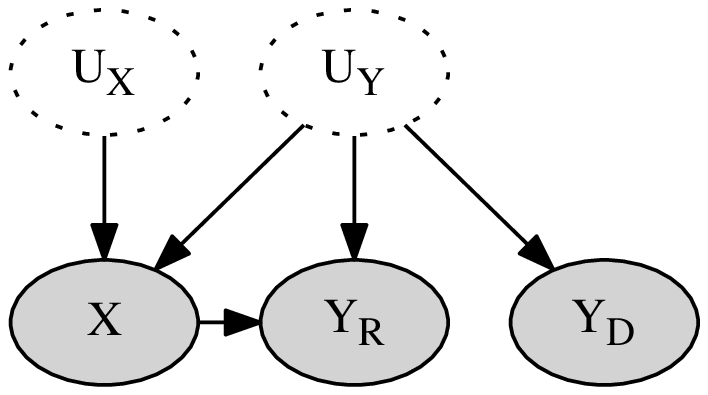}
  \end{minipage} \newline \textbf{(c)} \raggedright{Split-door criterion} &
      \vspace{-0.5cm}  Analyze subset of data where the auxiliary outcome $Y_D$ is independent of the treatment. & \vspace{-0.5cm} $Y_D \centernot\indep U_Y$  &  \vspace{-0.5cm}  Requires dependency between an auxiliary outcome and all confounders. & \vspace{-0.5cm} Measure marginal click-throughs on all pairs of products that have uncorrelated direct traffic. \\ 
       \bottomrule
      \end{tabular}
  }
      \captionof{figure}{Comparison of methods for estimating the effect of a treatment $X$ on an outcome $Y$. $W$ and $U$ represent all observed and unobserved confounders, respectively, that commonly cause both $X$ and $Y$.}
      \label{tab:compare-models}
\end{table}

\subsection{Instrumental Variables}
Both the split-door criterion and instrumental variable (IV) methods can be used to exploit naturally occurring variation in subsets of observational data to identify causal effects.
Importantly, however, they make different assumptions.
In IV methods, one uses an auxiliary variable $Z$, called an instrument, that is assumed to be exogenous and that systematically shifts the distribution of the cause $X$.
The validity of an instrument relies on two additional assumptions: first that it is effectively random with regard to potential confounders ($Z \indep U$), and second that the instrument affects the outcome $Y$ only through the cause $X$ ($Z \indep Y | X, U$).
Both of these conditions involve independence claims between observed and unobserved variables, making them impossible to test in practice~\citep{dunning2012natural}.

The split-door criterion also relies on an auxiliary variable, but one that relates to the outcome instead of the treatment.
Specifically, it exploits an auxiliary outcome $Y_D$ that serves as a proxy for unobserved common causes $U_Y$ under three important assumptions.
The first is that the cause $X$ does not affect $Y_D$ directly.
The second assumption requires that all unobserved confounders (between the cause and outcome) that affect $Y_R$  also affect $Y_D$. 
As with IV methods above, these two assumptions involve knowledge of an unobserved variable and, as a result, cannot be tested.
The third assumption requires independence between the cause $X$ and the auxiliary outcome $Y_D$.
Since both of these variables are observed, this assumption can be tested empirically so long as we are in the standard setting where statistical independence implies causal independence (\textit{Assumption 2}), equivalent to the assumption of \textit{faithfulness}~\citep{spirtes2000causation}.

It is difficult to compare these two sets of assumptions in general, but in different scenarios, one of these methods may be more suitable than the other. 
If a valid instrument is known to exist, for instance through changes in weather or as a result of a lottery, the variation it produces can and should be exploited to identify causal effects of interest.
The split-door criterion, in contrast, is most useful when one suspects there is random variation in the data, but cannot identify its source {\it a priori}. In particular, it is well-suited for large-scale data where the first two assumptions mentioned above are plausible, such as in digital or online systems.

\subsection{Back-door criterion}
\label{sec:splitdoor-regression}
Alternatively, the split-door criterion can be interpreted as using $Y_D$ as a proxy for all confounders $U_Y$, and estimating the causal effect whenever $Y_D$ (and hence $U_Y$) is independent of $X$.
Viewed this way, the split-door approach may appear to be nothing more than a variant of the back-door criterion where one conditions on $Y_D$ instead of $U_Y$, however there are two key differences between the two methods. 

First, substituting $Y_D$ for $U_Y$ in the  back-door criterion assumes that $Y_D$ is a perfect proxy for $U_Y$. This is a much stronger assumption than requiring that $Y_D$ be simply affected by $U_Y$, because any difference (e.g., measurement error) between $Y_D$ and $U_Y$ can invalidate the back-door criterion~\citep{spirtes2000causation}.
Second, the two methods differ in their approach to identification. The split-door criterion \textit{controls} for the effect of unobserved confounders by finding subsets of data where $X$ is not affected by $U_Y$, whereas the back-door criterion \textit{conditions} on a proxy for $U_Y$ to nullify the effect of unobserved confounders. Therefore, by directly controlling at the time of data selection, the split-door criterion focuses on admitting a subset of the data for analysis and simplifies effect estimation, whereas methods based on back-door criterion such as regression, matching, and stratification process the whole dataset and extract estimates via statistical models~\citep{morgan2014counterfactuals}.

To illustrate these differences, we compare mathematical forms of the split-door and back-door criteria in terms of regression equations. Conditioning on $Y_D$ using regression will lead to the following equation
\begin{equation}
y_r  = \rho'' x + \beta y_d + \epsilon''_{yr}, \nonumber
\end{equation}
applied to the entire dataset.
In contrast the split-door criterion leads to the simpler equation (as shown earlier in Section~\ref{sec:splitdoor-struc-eqns})
\begin{equation}
 y_r  = \rho x + \epsilon' _{yr}, \nonumber
\end{equation}
applied only to subsets of data where $X$ and $Y_D$ are independent.

\subsection{Methods based on empirical independence tests}
Finally, the split-door criterion is similar to recent work that proposes a data-driven method for determining the appropriate window size in regression discontinuity designs~\citep{cattaneo2015randomization,cattaneo2017comparing}.
In regression discontinuities, treatment (e.g., acceptance into a program) is assigned based on whether an observed variable (e.g., a test score) is above or below a pre-determined cutoff.
The assumption is that one can compare outcomes for those just above and just below the cutoff to estimate causal effects, but the central problem is how
far from the cutoff this assumption holds.
The authors present a data-driven method for selecting a window by testing for independence between the treatment and pre-determined covariates that are uncoupled to the outcome of interest.
This approach resembles the split-door criterion in that both use independence tests to determine which subsets of the data to include when making a causal estimate.
As a result, both methods are subject to concerns around multiple hypothesis testing, although the regression discontinuity setting typically involves many fewer comparisons than the split-door criterion (dozens instead of the thousands we analyze here) and occurs over nested windows.
For these reasons we treat multiple comparisons differently, estimating the error rate in identifying independent instances instead of adjusting nominal thresholds to try to eliminate errors.


%% file: amazonrecs_application.tex
\section{Application: Impact of a Recommender System}
\label{sec:amazon}
\begin{figure}[tb]
\center
\includegraphics[width=0.65\textwidth]{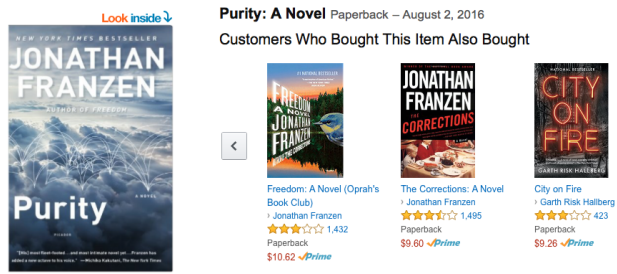}
\caption{Screenshot of a focal product, the book ``Purity'', and its recommendations on Amazon.com.}
\label{fig:recsys}
\end{figure}

We now apply the split-door criterion to the problem of estimating the causal impact of Amazon.com's recommender system.
Recommender systems have become ubiquitous in online settings, providing suggestions for what to buy, watch, read or do next~\citep{ricci2011}.
Figure~\ref{fig:recsys} shows an example of one of the millions of product pages on Amazon.com, where the main item listed on the page, or {\it focal product}, is the book ``Purity'' by Jonathan Franzen.
Listed alongside this item are a few {\it recommended products}---two written by Franzen and one by another author---suggested by Amazon as potentially of interest to a user looking for ``Purity''.
Generating and maintaining these recommendations takes considerable resources, and so a natural question one might ask is how exactly exposure to these recommended products changes consumer activity.

While simple to state, this question is difficult to answer because it requires an estimate of the counterfactual of what would have happened had someone visited a focal product but had not been exposed to any recommendations.
Specifically, we would like to know how much traffic recommender systems {\it cause}, over and above what would have happened in their absence.
Naively one could assume that users would not have viewed these other products without the recommender system, and as a result simply compute the observed click-through rate on recommendations~\citep{mulpuru2006,grau2009}.
As discussed earlier, however, this assumption ignores correlated demand: users might have found their way to some of these recommended products anyway via direct search or browsing, which we collectively refer to as ``direct traffic''.
For instance, some users who are interested in the book ``Purity'' might be fans of Franzen in general, and so might have directly searched on Amazon.com for his other works such as ``Freedom'' or ``The Corrections'', even if they had not been shown recommendations linking to them. 
The key to properly estimating the causal impact of the recommender, then, lies in accounting for this correlated demand between a focal product and its recommendations.

In this section we show how the split-door criterion can be used to eliminate the issue of correlated demand by automatically identifying and analyzing instances where demand for a product and one (or more) of its recommendations are independent over some time period $\tau$.
We do so by first formalizing this problem through a causal graphical model of recommender system traffic, revealing a structure amenable to the split-door criterion.
Then we apply the criterion to a large-scale dataset of web browsing activity on Amazon.com to discover thousands of instances satisfying the criterion. 
Our results show that a naive observational estimate of the impact of this recommender system overstates the causal impact on the products analyzed by a factor of at least two.
We conclude with a number of robustness checks and comments on the validity and generalizability of our results.

\subsection{Building the causal model}
\label{sec:amz-build-model}
The above discussion highlights that unobserved common demand for both a focal product and its recommendations can introduce bias in naive estimates of the causal click-through rate (CTR) on recommendations.
Referring back to Figure~\ref{fig:sdoor-model}, we formalize the problem as follows, with variables aggregated for each day:
\begin{itemize}
\item $X$ denotes the number of visits to the focal product $i$'s webpage.
\item $Y_{R}$ denotes recommendation visits, the number of visits to the recommended product $j$ through clicks on the recommendation for product $j$ on product $i$'s webpage. 
\item $Y_D$ denotes direct visits, the number of visits to product $j$ that did not occur through clicking on a recommendation.  These could be visits to $j$ from Amazon's search page or through direct visits to $j$'s webpage.
\item $U_Y$ denotes unobserved demand for product $j$, including both recommendation click-throughs and direct visits.
\item $U_X$ represents the part of unobserved demand for product $i$ that is independent of $U_{Y}$.
\end{itemize}

To apply the split-door criterion, we must investigate the plausibility of the \textit{connectedness} and \textit{independence} assumptions from Section~\ref{sec:splitdoor-graphical}.
First, the connectedness assumption states that both $Y_R$ and $Y_D$ are affected (possibly differently) by the same components of demand $U_Y$ for the product $j$.
As mentioned above, connectedness is especially plausible in the context of online recommender systems where products are easily reachable through multiple channels (e.g., search, direct navigation or recommendation click-through) and it is unlikely that demand for a product manifests itself exclusively through only one of these channels. Specifically, it is unlikely that there exists a component of demand for a product that manifests itself only through indirect recommendation click-throughs, but not through direct visits.
Put another way, for connectedness not to hold, it would have to be the case that users would have demand for a product \emph{only} if they arrived via a recommendation link, but not through other means.
To the best of our knowledge no path-specific feature of this sort exists on Amazon; thus, we expect the connectedness assumption to hold.

Second, with respect to the independence assumption, although we cannot rule out coincidental cancellation of effects that  result in $X \indep Y_D$ and violate the assumption, we expect such events to be unlikely over a large number of product pairs. Furthermore, for complementary product recommendations (which are the focus of this paper), we can logically rule out violation of the independence assumption because the demand for two complementary products are expected to be positively correlated with each other. 
Therefore,  it is reasonable to assume that the unobserved demand $U_{Y}$ (and all its sub-components) affect both $X$ and $Y_D$ in the same direction. For instance, let the effect of $U_Y$ be increasing for both $X$ and $Y_D$. Then the independence assumption is satisfied because the effect of $U_{Y}$ cannot be canceled out on the path $X \leftarrow U_{Y} \rightarrow Y_D$
if the effects of $U_{Y}$ (and any of its sub-components) on $X$ and $Y_D$ are all positive.
Given the above assumptions, the same reasoning from Section~\ref{sec:splitdoor-graphical} allows us to establish that $X \indep Y_D$ is a sufficient condition for causal identification.

\subsection{Browsing data}
Estimating the causal impact of Amazon.com's recommender system requires fine-grained data detailing activity on the site.
To obtain such information, we turn to anonymized browsing logs from users who installed the Bing Toolbar and consented to provide their anonymized browsing data through it.
These logs cover a period of nine months from September 2013 to May 2014 and contain a session identifier, an anonymous user identifier, and a time-stamped sequence of all non-secure URLs that the user visited in that session.
We restrict our attention to browsing sessions on Amazon.com, which leaves us with 23.4 million page visits by 2.1 million users spanning 1.3 million unique products.
Of these products, we examine those that receive a minimum of 10 page visits on at least one day in this time period,  resulting in roughly 22,000 focal products of interest.

Amazon shows many kinds of recommendations on its site.
We limit our analysis to the ``Customers who bought this also bought'' recommendations depicted in Figure~\ref{fig:recsys}, as these recommendations are the most common and are shown on product pages from all product categories.
To apply the split-door criterion, we need to identify focal product and recommended product pairs from the log data and separate out traffic for recommended products into direct ($Y_D$) and recommended ($Y_R$) visits.
Fortunately it happens to be the case that Amazon makes this identification possible by explicitly embedding this information in their URLs.
Specifically, given a URL for an Amazon.com page visit, we can use the \texttt{ref}, or referrer, parameter in the URL to determine if a user arrived at a page by clicking on a recommendation or by other means.
We then use the sequence of page visits in a session to identify focal and recommended product pairs by looking for focal product visits that precede recommendation visits. 
Further details about the toolbar dataset and construction of focal and recommended product pairs can be found in past work~\citep{sharma2015}.

\subsection{Applying the split-door criterion}
\label{sec:choose-tau}
Having argued for the assumptions underlying the split-door criterion and extracted the relevant data from browsing logs, the final step in estimating the causal effect of Amazon.com's recommendation system is to use the criterion to search for instances where a product and its recommendation have uncorrelated demand.

Recalling Section~\ref{sec:splitdoor-algorithm}, we employ a randomization test to search for 15-day time periods that fail to reject the null hypothesis that direct visits to a product and one (or more) of its recommended products are independent. The choice of 15 days represents a trade-off between two requirements: first, a time period large enough to yield reliable estimates; and second, a time period short enough that Amazon's recommendations for any given product are unlikely to have changed within that window.

The full application of the split-door criterion is as follows. For each focal product $i$ and each $\tau=15$ day time period:
\begin{enumerate}
 \item Compute $X^{(i)}$, the number of visits to the focal product on each day, and $Y_R^{(ij)}$, the number of click-throughs to each recommended product $j$. Also record the total direct visits $Y_D^{(j)}$ to each recommended product $j$.
 \item For each recommended product $j$, use the randomization test from Section~\ref{sec:choose-indep-test} to determine if $X^{(i)}$ is independent of $Y_D^{(j)}$ at a pre-specified significance level.\footnote{Here we filter out any time periods where $Y_D$ is exactly constant (because that will satisfy empirical independence conditions trivially).}
 \begin{itemize}
 \item  If $X^{(i)}$ is found to be independent of $Y_D^{(j)}$, compute the observed click-through rate (CTR), $\hat{\rho}_{ij\tau} = (\sum_{t=1}^{\tau}Y_R^{(ij)})/(\sum_{t=1}^{\tau}X^{(i)})$, as the causal estimate of the CTR. Otherwise ignore this product pair.
 \end{itemize}
 \item Aggregate the causal CTR estimate over all recommended products to compute the total causal CTR per focal product, $\hat{\rho}_{i\tau}$.
\end{enumerate}
Finally, average the causal CTR estimate over all time periods and focal products to arrive at the mean causal effect, $\hat{\rho}$, and compute the rate of erroneous split-door instances $\phi$ to estimate error in this estimate, as detailed in Appendices~\ref{sec:fdr} and \ref{sec:error-bar}.

\begin{figure}[tb]
\centering
\subfloat[Accepted at $\alpha$=0.95]{\includegraphics[width=0.47\textwidth]{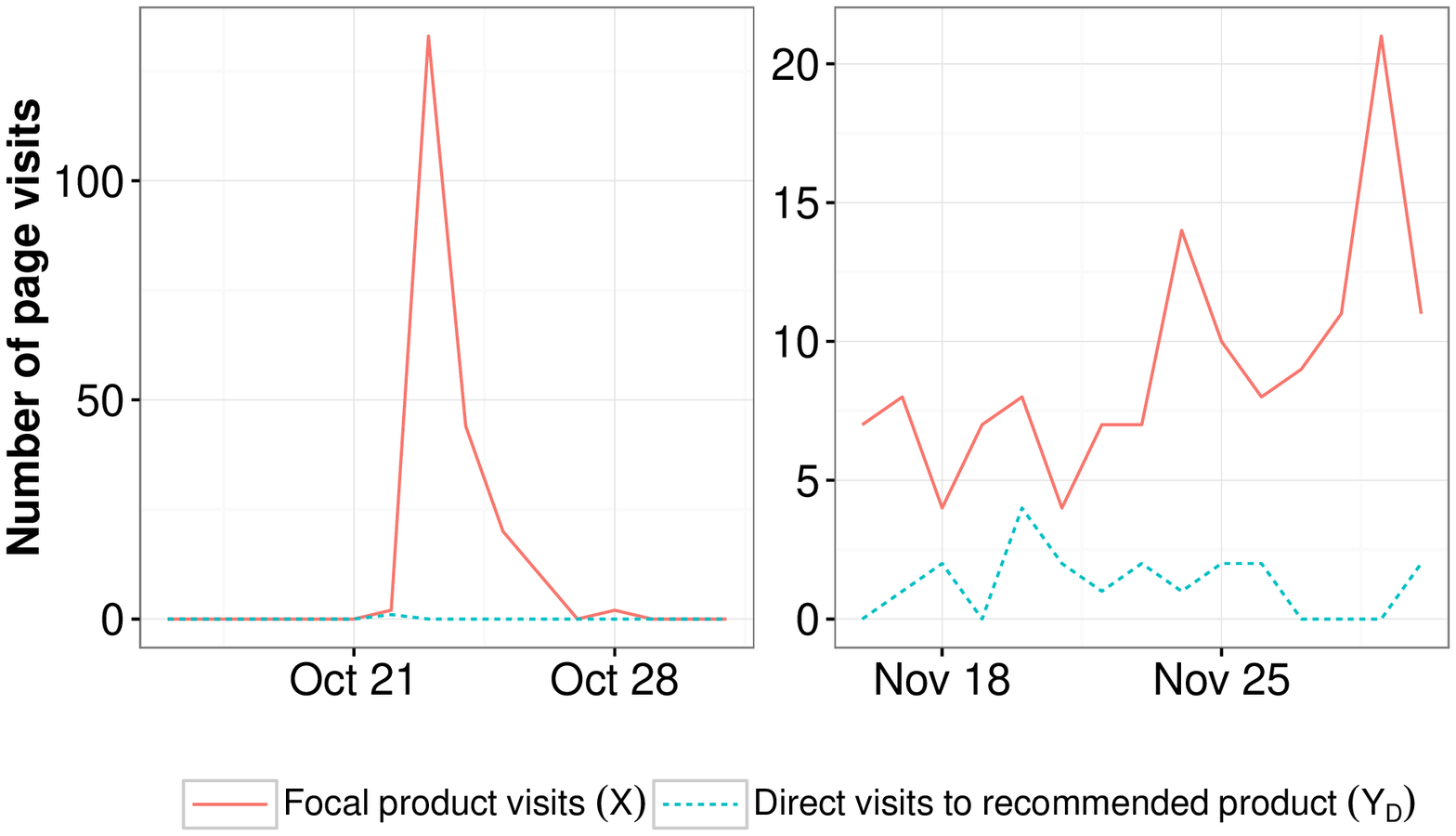}\label{fig:example-splitdoors}}%
\qquad
\subfloat[Rejected at $\alpha$=0.95]{\includegraphics[width=0.47\textwidth]{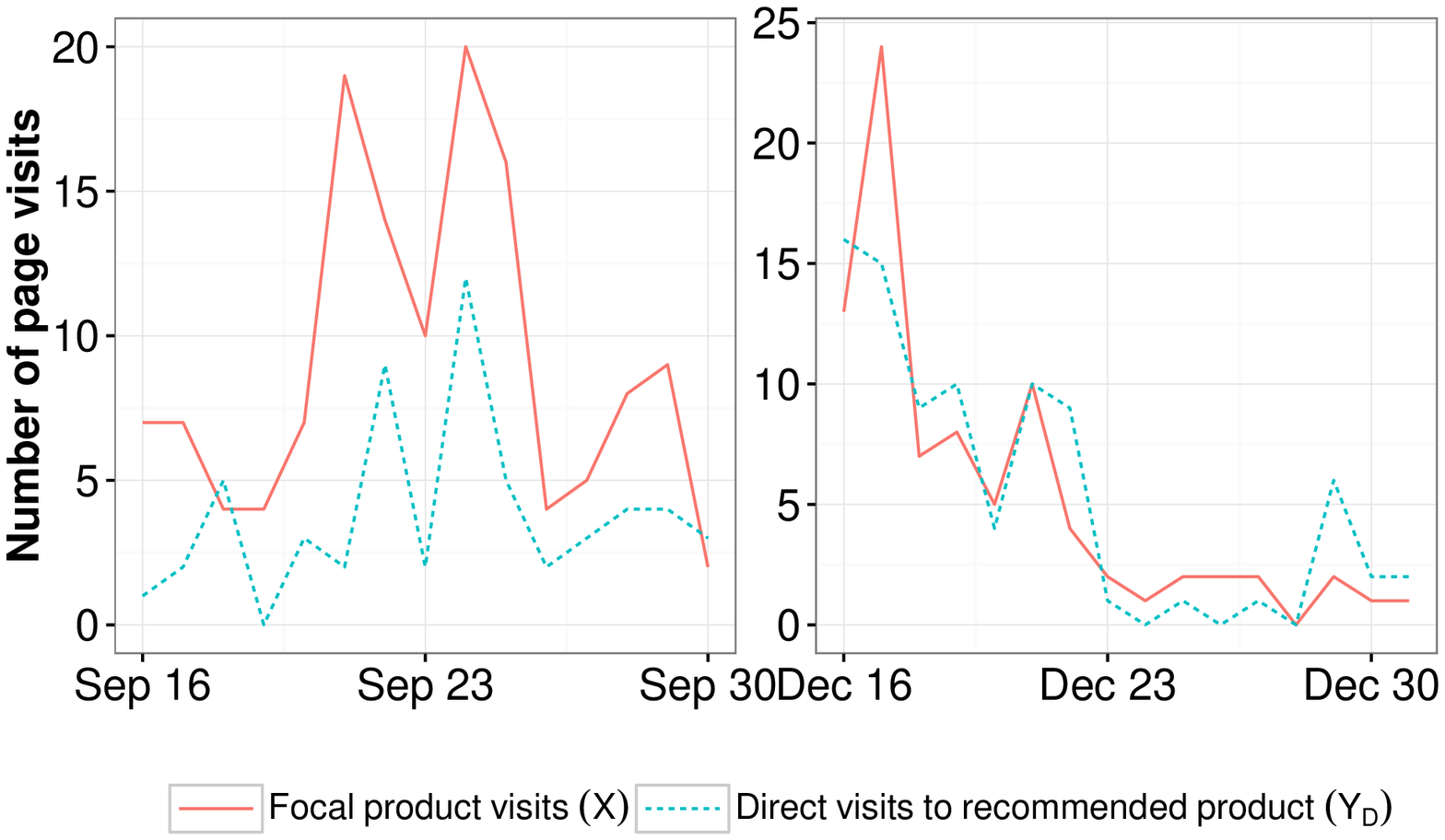}\label{fig:invalid-splitdoors}}
\caption{Examples of  time series for focal and recommended  products that are (a) accepted or (b) rejected by the split-door criterion at a significance level of $\alpha=0.95$ for the independence test.}
\end{figure}

\subsection{Results}
Applying the above algorithm results in over 114,000 potential split-door instances, where each instance consists of a pair of focal and recommended product over a 15-day time period. 
At a significance level of $\alpha = 0.95$,
we obtain more than 7,000 instances that satisfy the split-door criterion.
Consistent with previous work~\citep{sharma2015}, the corresponding causal CTR estimate $\hat{\rho}$ is 2.6\% (with the error bars spanning 2.0\% to 2.7\%), roughly one quarter of the naive observational estimate of $9.6\%$ arrived at by computing the click-through rate across all focal and recommended product pairs.
Put another way, these results imply that nearly 75\% of page visits generated via recommendation click-throughs would likely occur in the absence of recommendations.

Figure~\ref{fig:example-splitdoors} shows examples of product pairs that are accepted by the test at $\alpha = 0.95$.
The example on the left shows a focal product that receives a large and sudden shock in page visits, while direct visits to its recommended product remains relatively flat.
This is reminiscent of the examples analyzed in \cite{carmi2012} and \cite{sharma2015}.
The example on the right, however, shows more general patterns that are accepted under the split-door criterion but not considered by these previous approaches: although direct visits to both the focal and recommended products vary substantially, they do so independently, and so are still useful in our estimate of the recommender's effect.
Conversely, two example product pairs that are rejected by the test are shown in Figure~\ref{fig:invalid-splitdoors}. 
As is visually apparent, visit patterns for each of the focal and recommended product pairs are highly correlated, and therefore not useful in our analysis.

\begin{figure}[tb]
\centering
\subfloat[]{\includegraphics[width=0.31\textwidth]{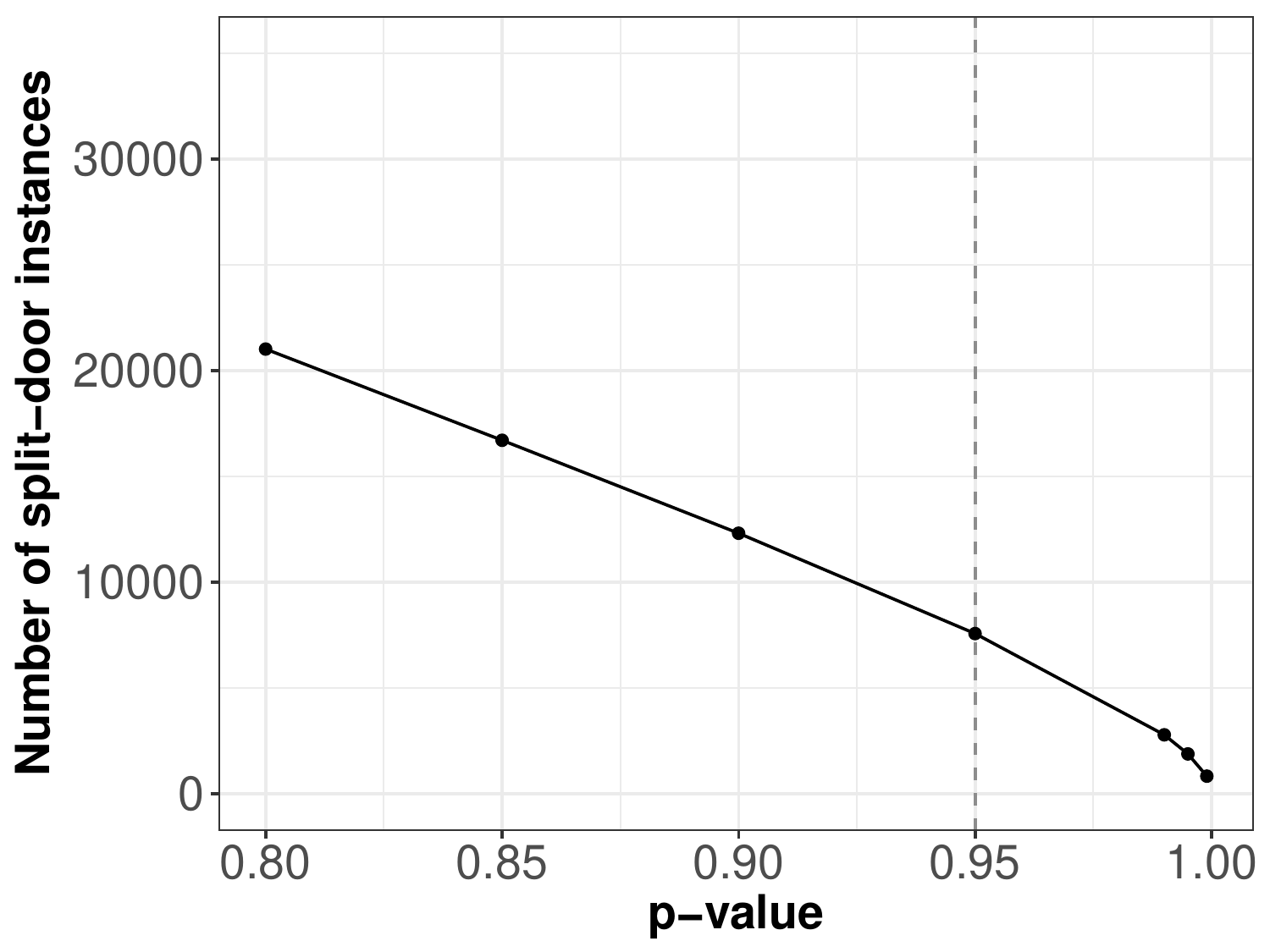}\label{fig:numexp-vs-pval}}%
\quad
\subfloat[]{\includegraphics[width=0.31\textwidth]{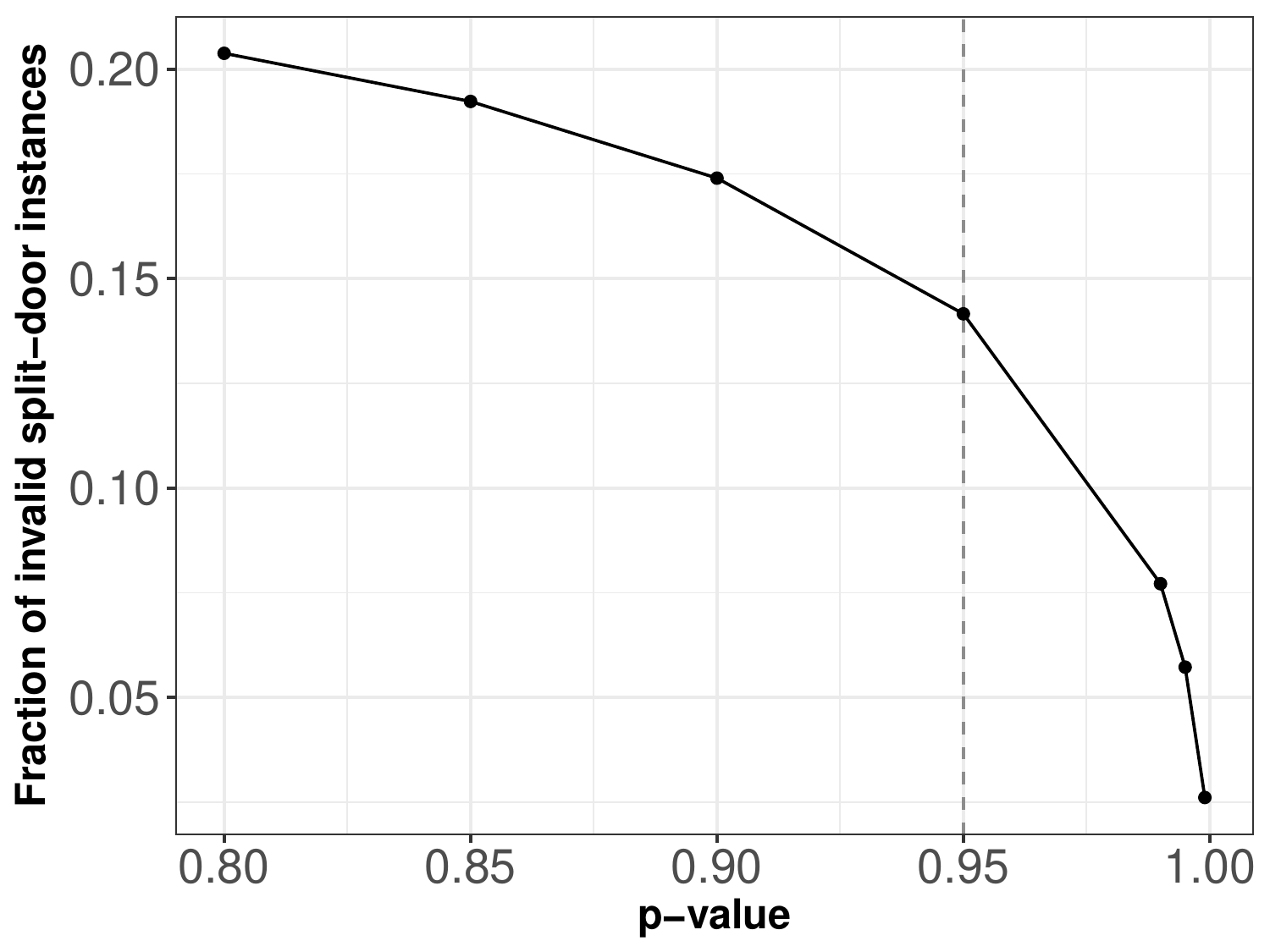}\label{fig:fdr-vs-pval}}%
\quad
\subfloat[]{\includegraphics[width=0.31\textwidth]{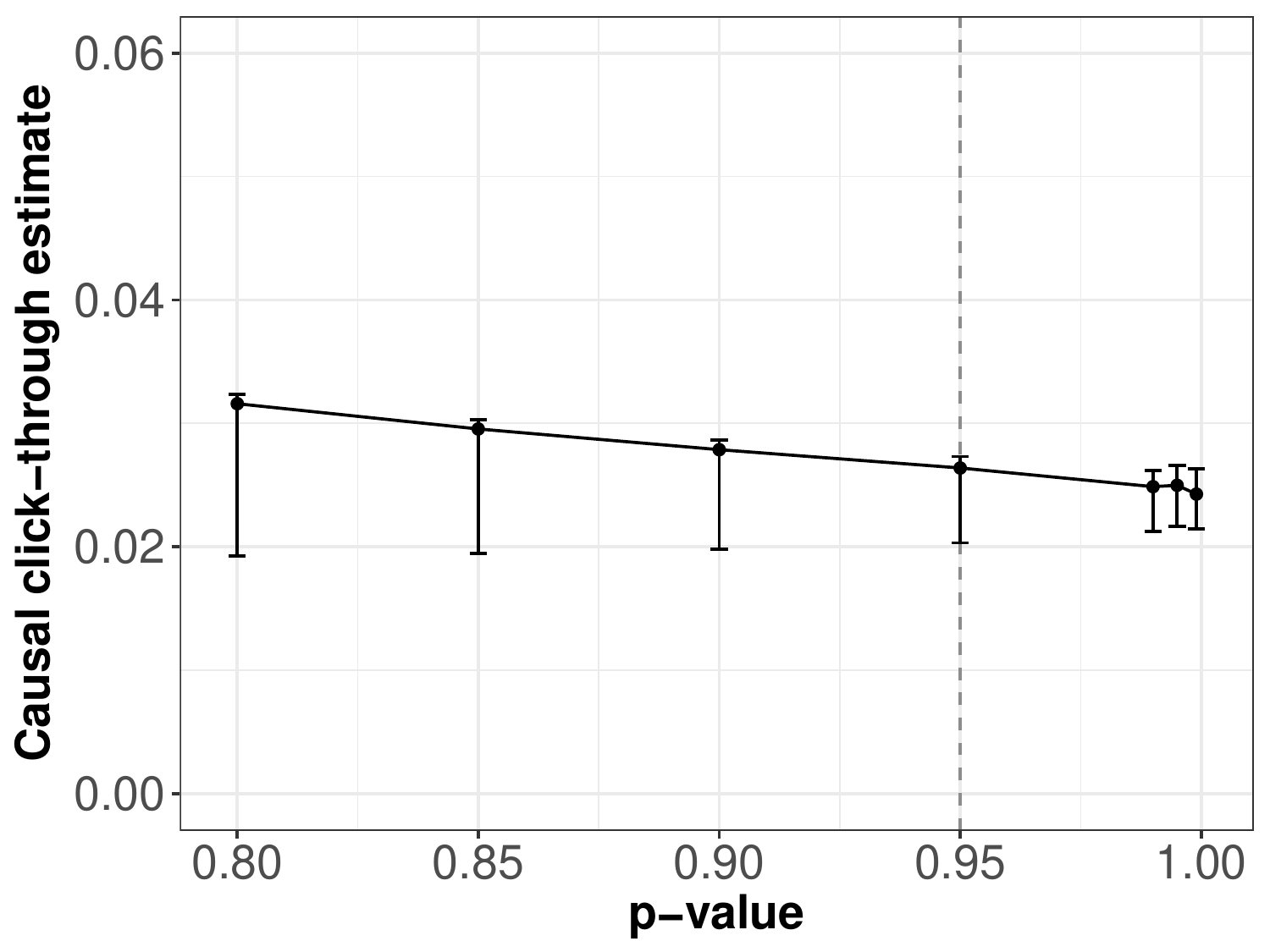}\label{fig:bounds-vs-pval}} 
\vspace{-1em}
\caption{Subplot (a) shows the number of valid split-door instances obtained as the p-value threshold ($\alpha$) is increased. Subplot (b) shows the expected fraction of erroneous instances ($\phi$) returned by the method for those values of $\alpha$. 
The corresponding estimate for causal CTR is shown in Subplot (c); error bars account for both bias due to $\phi$ and natural variance in the mean estimate.
}
\label{fig:varying-pval}
\end{figure}

Changing the nominal p-value threshold used in the independence test allows us to explore a tradeoff between
coverage across products in our dataset and the precision of our causal estimate.
As detailed in Appendix~\ref{sec:fdr}, a lower threshold results in more discovered instances, but with a higher likelihood of these instances being invalid.
For instance, Figures~\ref{fig:numexp-vs-pval} and \ref{fig:fdr-vs-pval}  show that decreasing the threshold to $\alpha = 0.80$ results in over 20,000 split-door instances covering nearly 11,000 unique focal products, but does so at the expense of increasing the expected fraction of invalid instances to 0.21, indicating that approximately one in five of the returned split-door instances may be invalid.
The result, summarized in Figure~\ref{fig:bounds-vs-pval}, is that the error bars on our estimate of $\rho$ increase as we decrease $\alpha$.
These error bars, calculated using Equation~\ref{eqn:bounds} from Appendix~\ref{sec:error-bar}, account for both bias due to erroneous split-door instances and the natural variance in the mean estimate due to sampling.\footnote{Note that the error bars are asymmetric; we expect erroneous split-door instances to drive the causal estimate up from its true value, under the assumption that demand for the two products are positively correlated with each other, as argued in Section~\ref{sec:amz-build-model}.}  
As $\alpha$ decreases, erroneous instances due to $\phi$ contribute to most of the magnitude of the error bars shown in Figure~\ref{fig:bounds-vs-pval}.
We observe that $\alpha=0.95$ offers a good compromise: error bounds are within $1$ percentage point and we obtain more than 7,000 split-door instances.

\begin{figure}[tb]
\center
\includegraphics[scale=0.6]{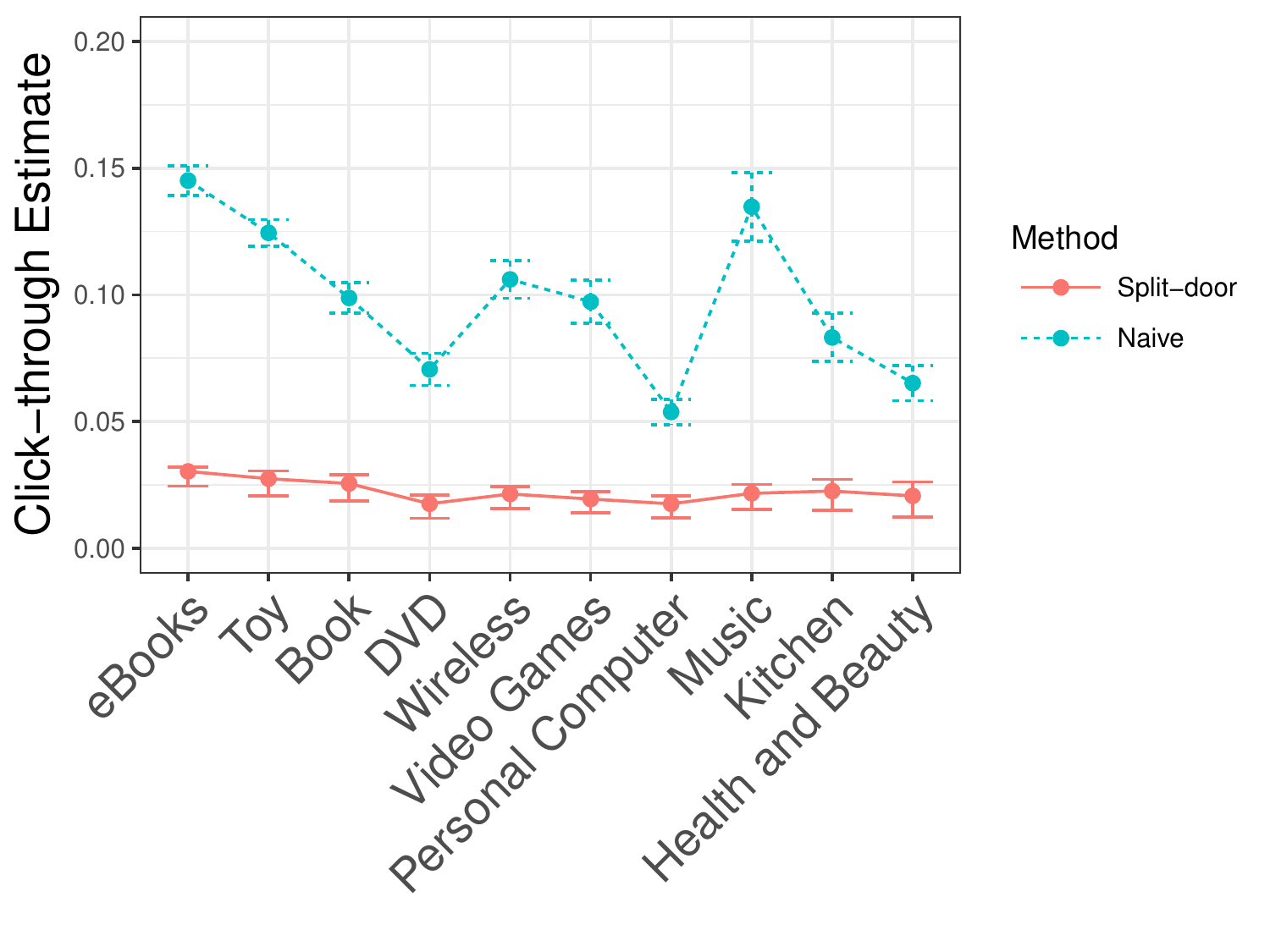} \vspace{-1em}
\caption{Comparison of the causal CTR with the naive observational CTR for products that satisfy the split-door criterion. Categories are ordered by the number of products found by the split-door criterion in each category, with \emph{eBooks} containing the most and \emph{Health and Beauty} the least.
}
\label{fig:rho-vs-pgroup}
\end{figure} 

Furthermore, we can break these estimates down by the different product categories present on Amazon.com.
Figure~\ref{fig:rho-vs-pgroup} shows the variation of $\hat{\rho}$ across the most popular categories, at a nominal significance level of $\alpha = 0.95$. For the set of focal products that satisfy the split-door criterion, we also compute the naive observational CTR.
We see substantial variation in the naive estimate, ranging from $14\%$ on \textit{e-Books} to $5\%$ on \textit{Personal Computer}. However, when we use the split-door criterion to compute estimates, we find that the causal CTR for all product categories lies below $5\%$.
These results indicate that naive observational estimates overstate the causal impact by anywhere from two- to five-fold across different product categories.

There are two clear advantages to the split-door criterion compared to past approaches for estimating the causal impact of recommender systems. First, we are able to study a larger fraction of products compared to instrumental variable approaches that depend on single-source variations \citep{carmi2012} or restricting our attention to mining only shocks in observational data~\citep{sharma2015}.
On the same dataset, the shock-based method in \cite{sharma2015} identified valid instances on 4,000 unique focal products, while the split-door criterion finds instances for over 5,000 unique focal products at $\alpha=0.95$, and over 11,000 at $\alpha=0.80$.
Second, the split-door criterion provides a principled method to select valid instances for analysis
by tuning $\alpha$, the desired significance level, while also allowing for an estimate of the fraction of falsely accepted instances, $\phi$.

\subsection{Threats to validity}
As with any observational analysis, our results rely on certain assumptions that may be violated in practice. Furthermore, results obtained on a subset of data may not be representative of the broader dataset of interest.
Here we conduct additional analyses to assess both the internal and external validity of our estimate of the causal effect of Amazon's recommendations.

\begin{figure*}[tb]
	\centering
    \subfloat[]{\includegraphics[width=0.47\textwidth]{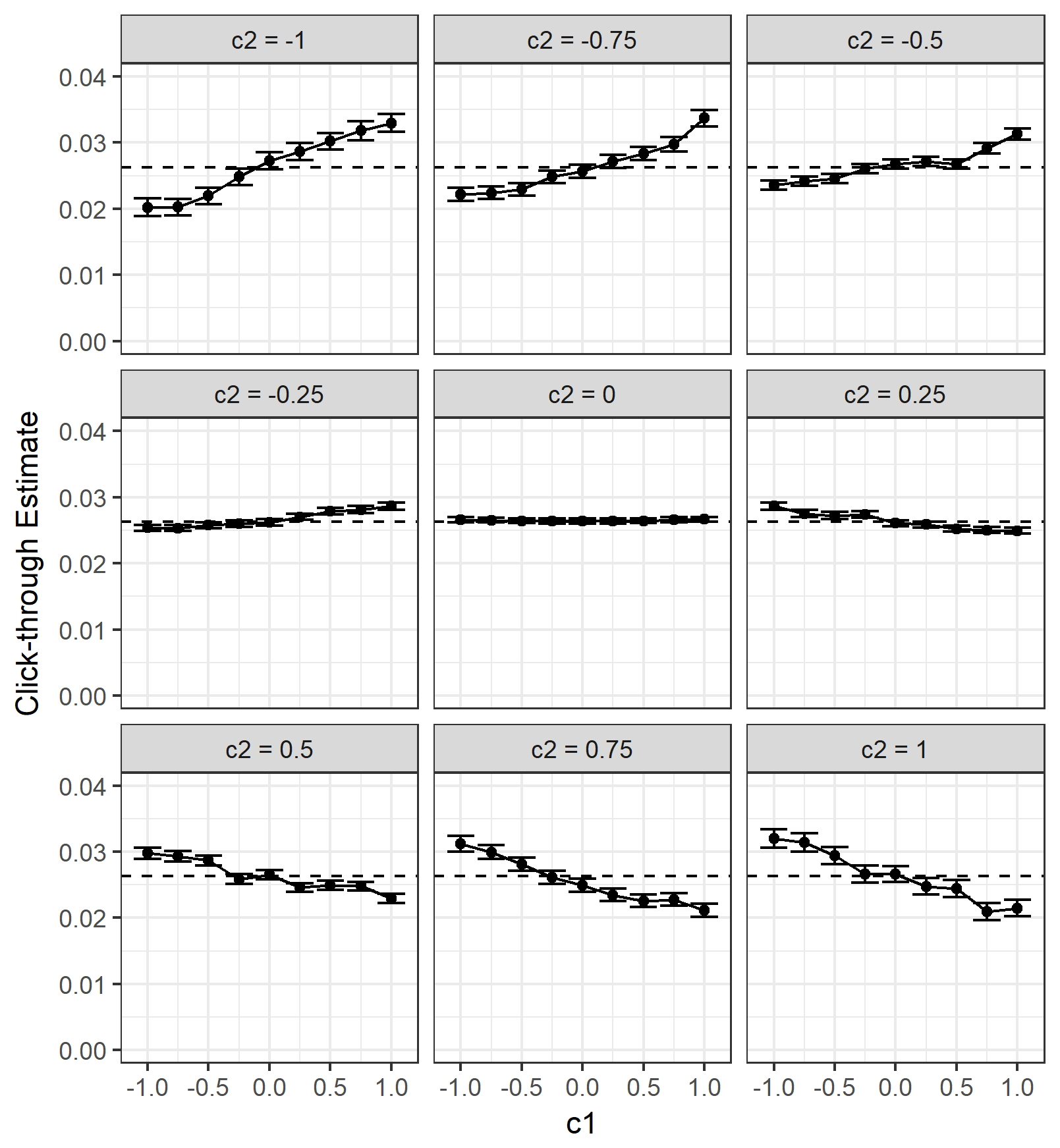}\label{fig:sens-res1}}%
	\qquad
	\subfloat[]{\includegraphics[width=0.47\textwidth]{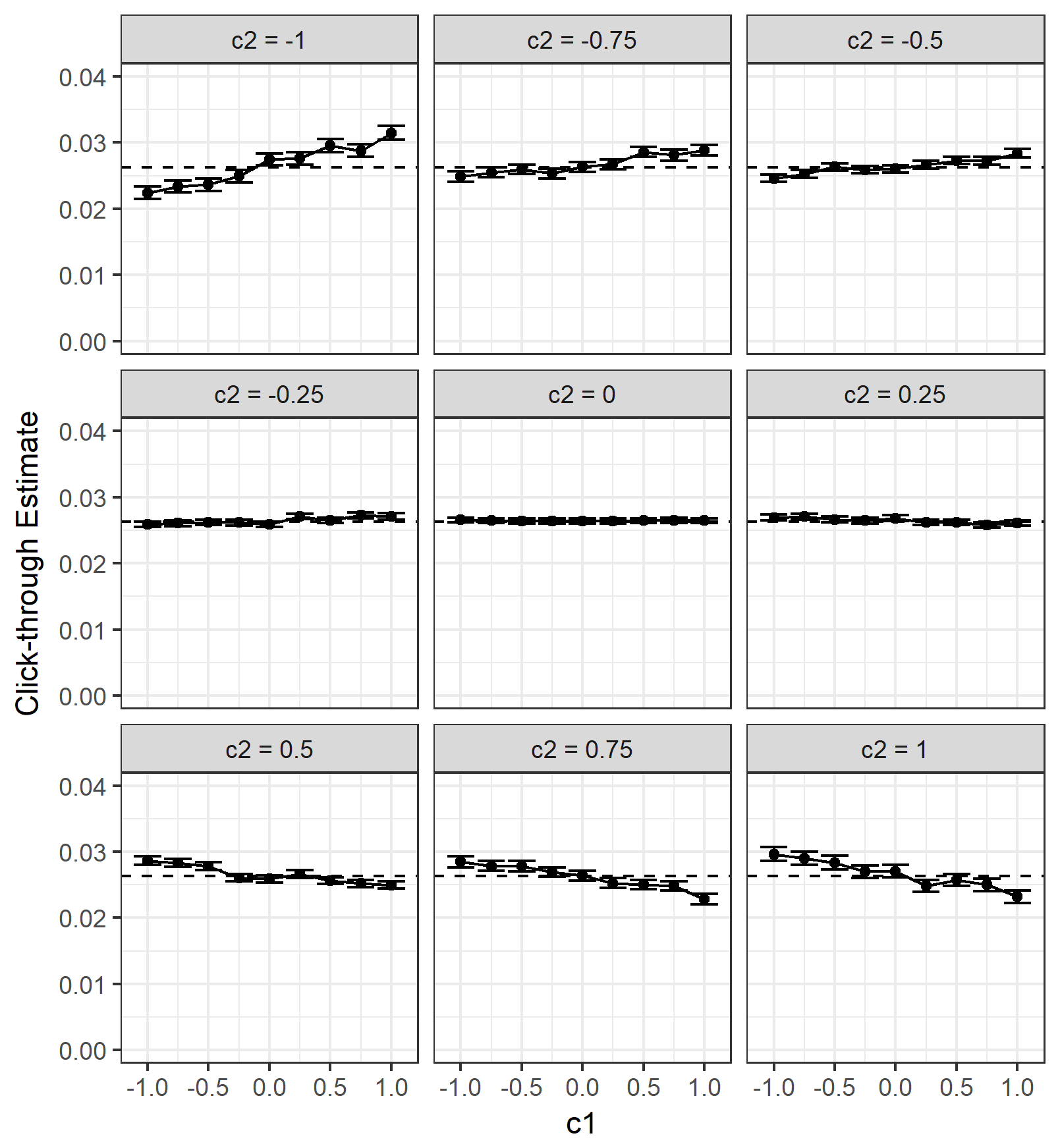}\label{fig:sens-res2}}%
    \vspace{-1em}
    \caption{Sensitivity analysis of the obtained click-through estimate. Panel (a) shows the scenario where all split-door instances may be invalid. Panel (b) assumes that at most half of the instances are invalid. The deviation of the true estimate from the obtained estimate increases as the magnitude of the confounding from $V_Y$ to $X$( $c_1$) or $Y$($c_2$) increases; however this deviation is lower in Panel (b). In both figures, dotted line shows the obtained CTR estimate.}
    \label{fig:sens-analysis}
\end{figure*}

\subsubsection{Internal validity: Sensitivity to the connectedness assumption}
\label{sec:amazon-sensitivity-analysis}
As described in Section~\ref{sec:sens-summary}, connectedness is the key identifying assumption for the split-door criterion. Here we describe a test for sensitivity of the obtained estimate ($\hat{\rho}$) to violations of the connectedness assumption. 

Referring to the causal model in Figure~\ref{fig:sens-connected1}, violation of the connectedness assumption implies that there exist components of unobserved demand $V_Y$ that affect both focal product visits $X$ and recommendation click-throughs $Y_R$, but not direct visits to the recommended product $Y_D$. For simplicity, let us assume that $V_Y$ is univariate normal and affects both $X$ and $Y_R$ linearly. We can write the corresponding structural equations for the causal model in Figure~\ref{fig:sens-connected2} for each split-door instance as
\begin{align}
    x &= c_1v_y + \epsilon_1 \\
    y_r &= f(x) + c_2v_y + \epsilon_2,
\end{align}
where $f$ is an unknown function, and  $\epsilon_1$ and $\epsilon_2$ are independent from all variables mentioned above and are also mutually independent. Note that $\epsilon_1$ includes the effect of $U_X$ and $\epsilon_2$ includes the effect of $U_Y$. For any split-door instance, the estimator from Section~\ref{sec:choose-tau} estimates the causal effect assuming that either $c_1$ or $c_2$ is zero.

To test the sensitivity of our estimate to the connectedness assumption, we take our actual data and introduce an artificial confound $V_Y$ by simulation, adding $c_1 V_Y$ to $X$ and $c_2 V_Y$ to $Y_R$, respectively, for a range of different $c_1$ and $c_2$ values.
We simulate $V_Y$ as a standard normal and vary $c_1$ and $c_2$ between $[-1,1]$, and compare these artificially confounded estimates to our actual estimate of $\hat{\rho} = 2.6\%$ for $\alpha=0.95$.
Figure~\ref{fig:sens-res1} shows the deviation between estimates using the actual and simulated data as $c_1$ and $c_2$ vary.  The difference is maximized when both $c_1$ and $c_2$ are high in magnitude and is negligible when either of $c_1$ or $c_2$ are zero.
These simulation results suggest a bilinear sensitivity to $c_1$ and $c_2$, a result we confirm theoretically in the case of a linear causal model in Appendix~\ref{sec:sensitivity}.

This analysis assumes that \emph{all} split-door instances violate the connectedness assumption.
Recognizing that this need not be the case, and that only some instances may be invalid, we introduce a third sensitivity parameter $\kappa$, which corresponds to the fraction of split-door instances that violate connectedness. For instance, we can test sensitivity of the estimate when at least half of the split-door instances satisfy connectedness, as done by \cite{kang2016invalidiv} for inference under multiple possibly invalid instrumental variables.
As shown in Figure~\ref{fig:sens-res2}, when $\kappa=0.5$ deviations from the obtained split-door estimate are nearly halved, resulting in more robust estimates. 

\begin{figure*}[tb]
	\centering
	\subfloat[]{\includegraphics[width=0.47\textwidth]{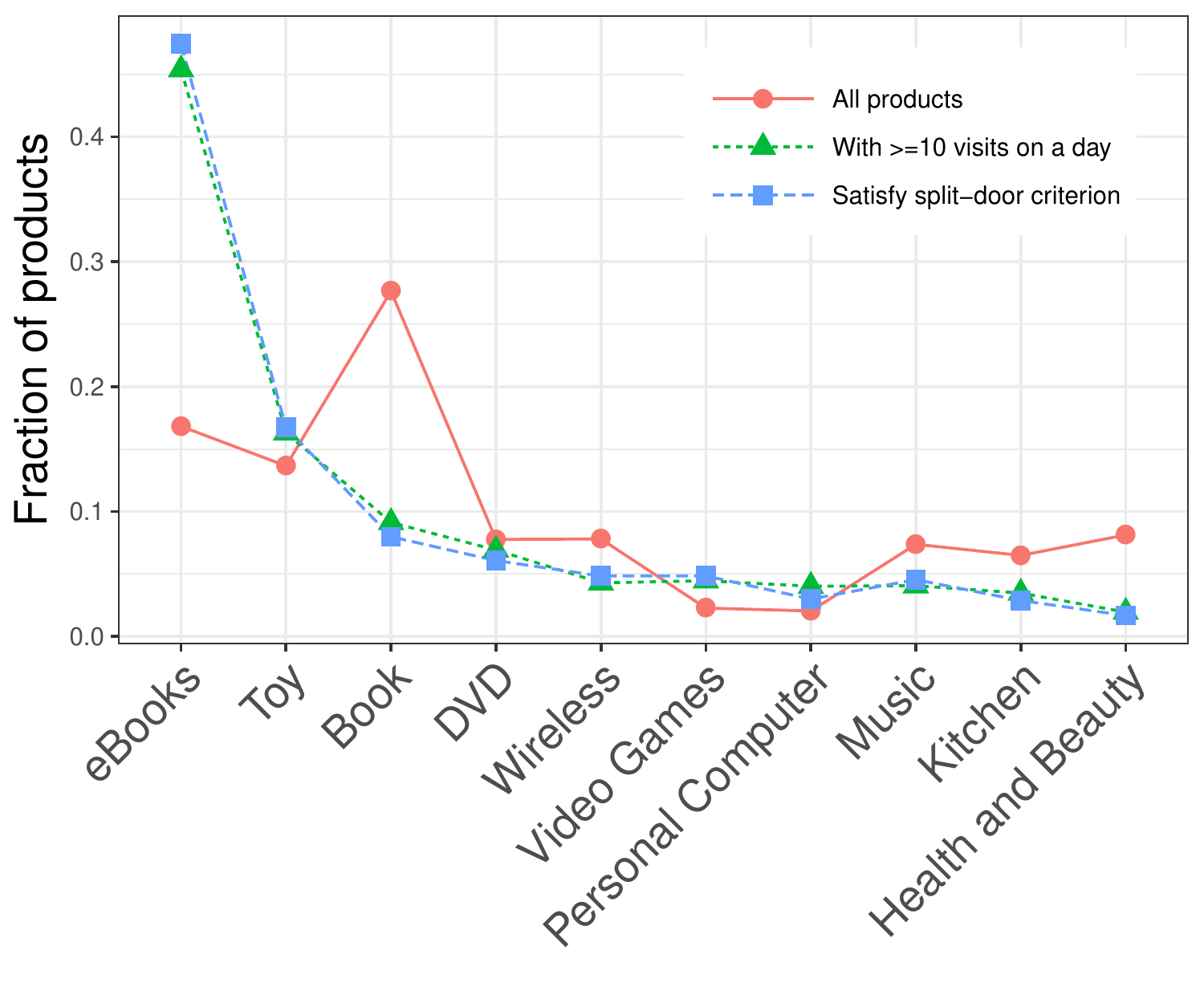}\label{fig:distr-by-pgroup-asins}}%
	\qquad
	\subfloat[]{\includegraphics[width=0.47\textwidth]{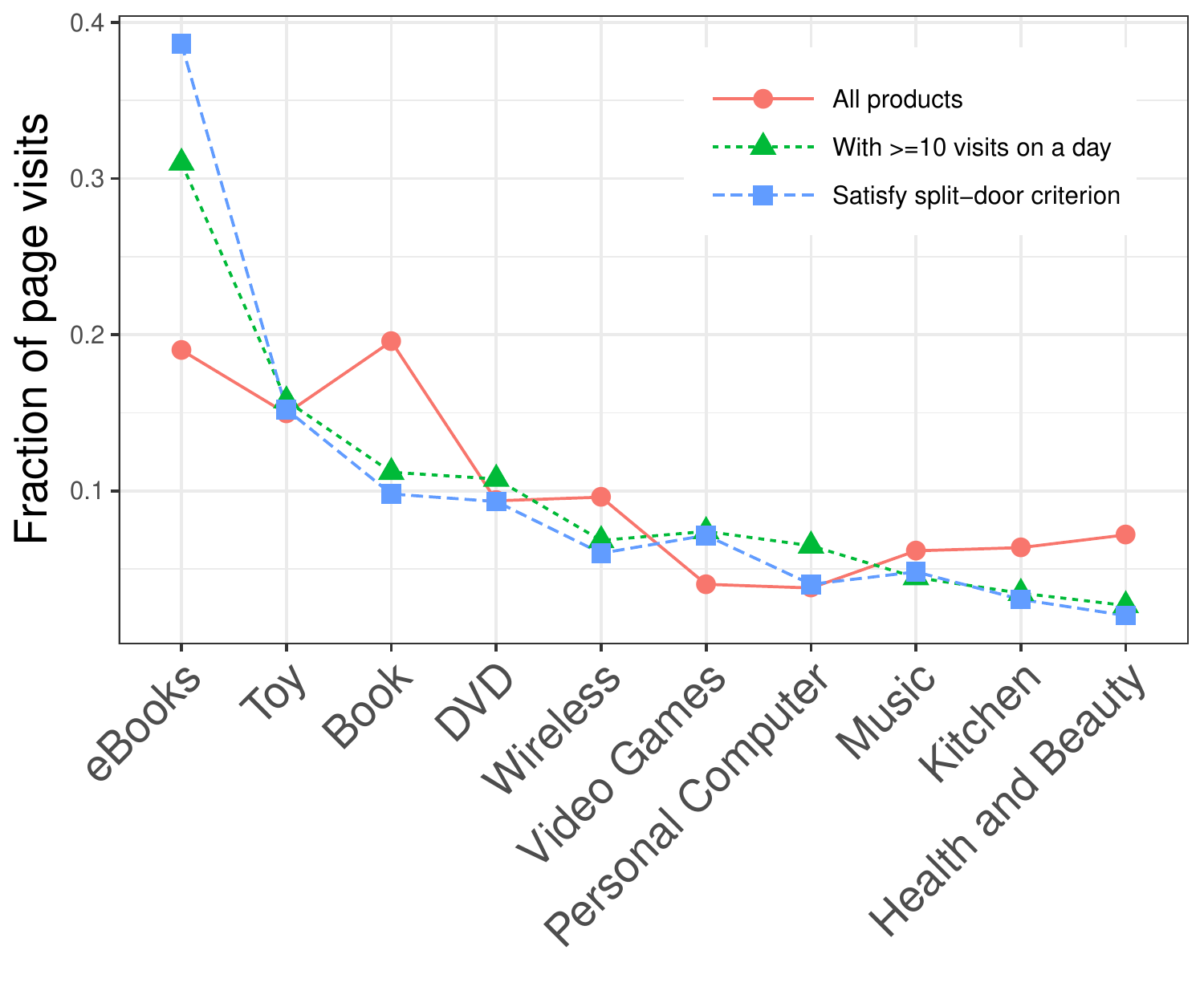}\label{fig:distr-by-pgroup-visits}}%

    \vspace{-1em}
	\caption{The distribution of products and total visits over product categories. Among products with at least 10 page visits on at least one day, the subset of focal products that satisfy the split-door criterion are nearly identical to the set of all products.  Fraction of page visits to those focal products show more variation, but the overall distributions are similar.}
        \label{fig:distr-by-pgroupi-visits}
\end{figure*}

\subsubsection{External validity: Generalizability} Although the split-door criterion yields valid estimates of the causal impact of recommendations for the time periods where product pairs are found to be statistically independent, it is important to emphasize that products in the split-door sample may not be selected at random, thus violating the \textit{as-if-random}~\citep{angrist1996} assumption powering generalizability for natural experiments.
As a result, care must be taken to extrapolate these estimates to all products on Amazon.com.

Fortunately, as shown in Figure~\ref{fig:distr-by-pgroupi-visits}, the distribution of products and page visits in our sample closely matches the inventory and activity on Amazon.com.
Products with at least one valid split-door time period span many product categories and cover nearly a quarter of all focal products in the dataset at $\alpha=0.95$.
Figure~\ref{fig:distr-by-pgroup-asins} shows that the distribution of products analyzed by the split-door criterion across different product categories is almost identical to the overall set of products. Figure~\ref{fig:distr-by-pgroup-visits} shows a similar result for the number of page visits of these products across different product categories, except for eBooks which are over-represented in valid split-door instances. 
For comparison, we apply the same popularity filter that we used for the split-door criterion---at least 10 page visits on at least one day---to the dataset with all products.

Although these results do not necessarily imply that the as-if-random assumption is satisfied (indeed it is very likely not satisfied) they do indicate that the split-door criterion at least allows us to estimate causal effects over a diverse sample of popular product categories, which is a clear improvement over past work~\citep{carmi2012,sharma2015}.


%% file: discussion.tex
\section{Discussion}
\label{sec:discussion}

In this paper we have presented a method for computing the causal effect of a variable $X$ on another variable $Y_R$  whenever we have an additional variable $Y_D$ which follows some testable conditions, and have shown its application in estimating the causal impact of a recommender system. We now
suggest guidelines to ensure proper use of the criterion and discuss other applications for which it might be used.

\subsection{Guidelines for using the criterion}
As with any non-experimental method for causal inference, the split-door criterion rests on various untestable assumptions and requires making certain modeling choices. We encourage researchers to reason carefully about these assumptions, explore sensitivity to modeling choices, and examine threats to the validity of their results.

\subsubsection{Reason about assumptions}
The split-door criterion relies on two untestable assumptions: \textit{independence} (of $X$ and $Y_D$), and \textit{connectedness} (i.e. non-zero causal effect of $U_Y$ on $Y_D$).
The independence assumption is a standard assumption for observational causal inference. Barring coincidental equality of parameters such that the effect of unobserved confounders on $X$ and $Y_D$ cancel out, the independence assumption is likely to be satisfied. Nonetheless we encourage researchers to think carefully about this assumption in applying the criterion in other domains. Depending on the application it may be possible to rule out such cancellations. For example, in our recommendation system study we expect demand for the focal and recommended product to be correlated. Therefore, the causal effect of demand on both products is expected to be directionally identical, and hence cancellation becomes impossible.

The connectedness assumption is potentially more restrictive.
In general, it is plausible whenever measurements $Y_R$ and $Y_D$ are additive components of the same tangible outcome $Y$ that can be reached by similar means.
That said, connectedness remains an untestable assumption where, once again, domain knowledge should be used to assess its plausibility.
For instance, even when $Y_R$ and $Y_D$ are additive components, in some isolated cases, $U_{Y}$ may not be connected to $Y_D$ at all.
In a recommender system this can happen when customers with pre-existing interest in a product somehow visit it only through recommendation click-throughs from other products. 
In such a scenario, the split-door criterion would be invalid.
We note, however, that this situation can arise only in the (unlikely) event that no such user found the product directly. 
When there is even a small number of users that visit the product directly, the split-door criterion will again be valid and, depending on the precision of the statistical independence condition, can be applied.

\subsubsection{Explore sensitivity to test parameters}
A key advantage of the split-door criterion is that once these two assumptions are met, it reduces the problem of causal identification
to that of implementing a test for statistical independence.
At the same time, this requires choosing a suitable statistical test and deciding on any free parameters the test may have.
For instance, in the case of the randomization test used here, there is a significance level $\alpha$ used to determine when to accept or reject focal and recommended product pairs as statistically independent.
Any such parameters should be varied to check the sensitivity of estimates to these choices, as in Figures \ref{fig:fdr-vs-pval} and \ref{fig:bounds-vs-pval}.

\subsubsection{Examine threats to validity}
After identifying and estimating the effect of interest, one should examine both the internal and external validity of the resulting estimate.
In terms of internal validity, we recommend conducting a sensitivity analysis to assess how results change when the assumptions required for identification are violated.
In the case of the recommender system example, we simulated violations of the connectedness assumption by artificially adding correlated noise to $X$ and $Y_R$ (but not $Y_D$) and re-ran the split-door method to look at variation in results, as shown in Figure~\ref{fig:sens-analysis}.

Finally, after establishing internal validity, one needs to consider how useful the resulting estimate is for practical applications. As remarked earlier and demonstrated in our recommender system application, the split-door criterion is capable of capturing the local average causal effect for a large sample of the dataset that satisfies the required independence assumption ($X \indep Y_D$). 
The argument has been made that such local estimates are indeed useful in themselves~\citep{imbens2010better}.
That said, the sample may not be representative of the entire population, and so one must always be careful to qualify an extension of the split-door estimate to the general population.
Naturally, the more instances discovered by the method, the more likely the estimate is to be of general use.
Additionally, we recommend that researchers perform checks similar to those in Figure~\ref{fig:distr-by-pgroupi-visits} to compare the distribution of any available covariates to check for differences between the general population and instances that pass the split-door criterion.

\subsection{Potential applications of the split-door criterion}
The key requirement of the split-door criterion is that the outcome variable must comprise two distinct components: one that is potentially affected by the cause, and another that is not directly affected by it.
In addition, we should have sufficient reason to believe that the two outcome components share common causes (i.e. the connectedness assumption must be satisfied), and that one of outcome variables can be shown to be independent of the cause variable (i.e. the independence assumption must be satisfied). 
These might seem like overly restrictive assumptions that limit applicability of the criterion, but in this section we argue that there are in fact many interesting cases where the split-door criterion can be employed.

As we have already noted, recommendation systems such as Amazon's are especially well-suited to these conditions, in large part because $Y_D$ has a natural interpretation of ``direct traffic", or any traffic that is not caused by a particular recommendation.
Likewise the criterion can be easily applied to other online systems that automatically log user visits, such as in estimating the causal effect of advertisements on search engines or websites.
Somewhat more broadly, time series data in general may be amenable to the split-door criterion, in part because different components of the outcome occurring at the same time are more likely to be correlated than components that share other characteristics, and in part because time series naturally generate many observations on the input and output variables, which permits convenient testing for independence.

For example, consider the problem of estimating the effect of social media on news consumption.  There has been recent interest \citep{flaxman2016filter} in how social media websites such as Facebook impact the news that people read, especially through algorithmic recommendations such as those for ``Trending news". Given time series data for user activity on a social media website and article visits from news website logs, we can use the split-door criterion to estimate the effect of social media on news reading. Here $Y_R$ would correspond to the visits that are referred from social media, and $Y_D$ would be all other direct visits to the news article. Most websites record the source of each page visit, so obtaining these two components for the outcome---visits to an article through social media and through other means---should be straightforward. Whenever people's social media usage is not correlated with direct visits to a news article, we can identify the causal effect of social media on news consumption. Similar analysis can be applied to problems such as estimating the effect of online popularity of politicians on campaign financing or the effect of television advertisements on purchases. 

Finally, although we have focused on online settings for which highly granular time series data is often collected by default, we note that there is nothing intrinsic to the split-door criterion that prevents it from being applied offline. For example, many retailers routinely send direct mail advertisements to existing customers whom they identify through loyalty programs. The split-door criterion could easily be used to estimate the causal effect of these advertisements on product purchases: $X$ would be the number of customers that are sent an advertisement; $Y_R$ would be the customers among them who purchased the product; and $Y_D$ would be the number of customers who bought the product without receiving the mailer. 
More generally, the split-door criterion could be used in any context where the outcome of interest can be differentiated into more than one channel.

\section{Conclusion}
\label{sec:conc}
In closing we note that the split-door criterion is just one example of a more general class of methods that adopt a data-driven approach to causal discovery~\citep{jensen2008automatic,sharma2015,cattaneo2015randomization,grosse2016identification}. As we have discussed, data-driven methods have important advantages over traditional methods for exploiting natural variation---allowing inference to be performed on much larger and more representative samples---while also being less susceptible to unobserved confounders than back-door identification strategies. As the volume and variety of fine-grained data continues to grow, we expect these methods to increase in popularity and to raise numerous questions regarding their theoretical foundations and practical applicability.


%% file: appendix.tex
\appendix

\section{Estimating the fraction of erroneous split-door instances}\label{sec:fdr}

Let the expected fraction of erroneous $X$-$Y_D$ pairs---split-door instances---returned by the method be $\phi$. In the terminology of multiple testing, $\phi$ refers to the \textit{False Non-Discovery Rate (FNDR)} \citep{delongchamp2004multiple}. This is different from the more commonly used False Discovery Rate (FDR) \citep{farcomeni2008fdrreview},  since we deviate from standard hypothesis testing by looking for split-door instances that have a p-value higher than a pre-determined threshold.  Given $m$ hypothesis tests and a significance level of $\alpha$, we show that the false non-discovery rate $\phi$ for the split-door criterion can be characterized as 
\begin{equation} \label{eqn:fdr}
    \phi_\alpha \leq \frac{(1-\alpha)\pi_{dep}m}{W_{\alpha}},
\end{equation}
where $\pi_{dep}$ is the fraction of actually dependent $X$-$Y_D$ instances in the dataset and $W_\alpha$ is the observed number of $X$-$Y_D$ instances returned by the method at level $\alpha$.

The above estimate can be derived using the framework proposed by \cite{storey2002direct} under two assumptions. The first is that the that the distribution of p-values under the null hypothesis is uniform, and the second is that the distribution of p-values under the alternative hypothesis is stochastic smaller than the uniform distribution.
 Let the number of invalid instances found using the split-door criterion be $T$. Then, by definition, the false non-discovery rate can be written as: 
$$\phi_\alpha = \operatorname{E}\bigg[\frac{T}{W}\bigg|W>0\bigg]\,.$$

Since the alternative distribution is stochastically smaller than uniform, we can arrive at an upper bound by replacing $T$ by the expected number of split-door instances if the alternative distribution were uniform, $(1-\alpha)*m_{dependent} = (1-\alpha)*\pi_{dep}*m$, giving
\begin{equation}
    \phi_\alpha \leq \frac{(1-\alpha)*\pi_{dep}m}{W_{\alpha}}\,.
\end{equation}
Here $\pi_{dep}$ is unknown, so it needs to be estimated. 
A common approach is to estimate the fraction of actually independent instances or null hypotheses $\pi_{indep}$ and then use  $\pi_{dep}=1-\pi_{indep}$ \citep{delongchamp2004multiple}.
For robustness, we suggest using multiple procedures to estimate $\pi_{indep}$ and verify sensitivity of results to the choice of $\pi_{indep}$. In this paper, we  use two different estimates, derived from \cite{storey2003sam}, \cite{storey2002direct} (\textit{Storey's} estimate); and \cite{nettleton2006fdr}, \cite{liang2012fdr} (\textit{Nettleton's} estimate).

Storey's estimate is defined as
\begin{equation}
\hat{\pi}_{indep} =\frac{W_\lambda}{m(1-\lambda)},
\end{equation}
where $\lambda \in [0,1)$ is a tunable parameter---similar in interpretation to $\alpha$---and $W_\lambda$ is the number of hypothesis tests having a p-value higher than $\lambda$.
The choice of $\lambda$ involves a bias-variance tradeoff, with $\lambda=0.5$ being a common choice, as in the SAM software developed by \cite{storey2003sam}.

Nettleton's estimate, on the other hand,
chooses the effective value of $\lambda$ adaptively, based on the observed p-value distribution. First, the p-value distribution is summarized in a histogram containing $B$ bins. Then, a threshold $\lambda$ is chosen as the index ($I$) corresponding to the left-most bin whose count fails to exceed the average count of the bins to its right.  This results in the following estimate, where $\lambda=(I-1)/B$:
\begin{equation}
\hat{\pi}_{indep} = \frac{W_\lambda}{m(1-\lambda)} = \frac{W_\lambda}{m(1-\frac{I-1}{B})}\,. 
\end{equation}

Applying each of these to the $m=114,469$ focal and recommended product pairs analyzed in Section~\ref{sec:amazon} allows us to estimate the true number of dependent $X$-$Y_D$ pairs in the dataset, $\pi_{dep}$. At $\alpha=0.95$, both methods give very similar results ($\pi_{dep,Storey}=0.184$, $\pi_{dep,Nettleton}=0.187$); we use $\pi_{dep}=0.187$ in our analysis.

\section{Sensitivity analysis for the connectedness assumption}\label{sec:sensitivity}
In this section we analyze the sensitivity of an estimate obtained using the split-door criterion to violations of the connectedness assumption. As Figure~\ref{fig:sens-connected1} shows, violation implies that there exist variables $V_Y$ that affect only $X$ and $Y_R$ but not $Y_D$. We use the structural equation model from Section~\ref{sec:splitdoor-struc-eqns} to illustrate sensitivity analysis. 

Given that the unobserved confounders can be broken down into two components $U_Y$ and $V_Y$, we can rewrite the linear structural equations from Equation~\ref{eqn:linear-struc-eqns1} as:
\begin{align}
    x  &= \eta u_x + \gamma_1 u_y + c_1 v_y + \epsilon _x & & \\
    y_r  &= \rho x + \gamma_2 u_y + c_2 v_y + \epsilon _{yr} & & \\
    y_d  &= \gamma_3 u_y + \epsilon _{yd}, 
\end{align}
with two additional parameters $c_1$ and $c_2$ denoting the effect of the unobserved variable $V_Y$ on $X$ and $Y_R$, respectively.
Applying the split-door criterion $X \indep Y_D$, we write the following equations for each obtained split-door instance:
\begin{align}
    x &= \eta u_x + c_1 v_y + \epsilon' _x \\
    y_r &= \rho x + c_2 v_y + \epsilon' _{yr} 
\end{align}
Here $V_Y$ is unobserved and hence the causal effect is not identified. Using (B.5) as an estimating equation will lead to a biased estimate of the causal effect due to the confounding effect of the unobserved common cause $V_Y$. Note that this structure is identical to the omitted variable bias problem in back-door and conditioning-based methods \citep{harding2009sensitivity}. Consequently, we obtain a similar bilinear dependence of the split-door estimate to sensitivity parameters $c_1$ and $c_2$. 

Specifically, the split-door method regresses $Y_R$ on $X$ to obtain an estimate $\hat{\rho}$ for each obtained instance. When connectedness is violated, the bias of this estimate can be characterized as,
\begin{align*}
    \hat{\rho} = (X^TX)^{-1}X^TY_R &= \frac{\sum_i x_iy_{r_i}}{\sum_j x_j^2} \\
                                   &=\frac{\sum_i x_i(\rho x_i + c_2 v_{y_i} + \epsilon'_{yr_i}) }{\sum_j x_j^2}\\
                                   &=\rho + c_2\frac{\sum_i v_{y_i} x_i}{\sum_j x_j^2} + \frac{\sum_i \epsilon'_{yr_i} x_i}{\sum_j x_j^2} \,,
\end{align*}
where we use (B.5) to expand $y_{r_i}$. As in Section~\ref{sec:splitdoor-algorithm}, let $\tau$ denote the sample size for each split-door instance. When $X$ and $V_Y$ are both standardized to have zero mean and unit variance, and taking expectation on both sides, we obtain,
\begin{align}
    \operatorname{E}[\hat{\rho}] &= \rho + c_2\operatorname{E}[\frac{1}{\tau}\sum_i v_{y_i}x_i]+ \operatorname{E}[\frac{1}{\tau}\sum_i \epsilon'_{yr_i} x_i] \nonumber \\
                                 &= \rho + c_2\operatorname{E}[\frac{1}{\tau}\sum_i v_{y_i}(\eta u_{x_i} + c_1 v_{y_i} + \epsilon'_{x_i})] \nonumber \\
                                 &= \rho + c_1c_2\operatorname{E}[\frac{1}{\tau}\sum_i v_{y_i}v_{y_i}] +\eta\operatorname{E}[\frac{1}{\tau}\sum_i v_{y_i}u_{x_i}]+\operatorname{E}[\frac{1}{\tau}\sum_i v_{y_i} \epsilon'_{x_i}] \nonumber \\
     \operatorname{E}[\hat{\rho}]  &= \rho + c_1c_2
\end{align}
where we use the independence of error terms and that $U_X \indep V_Y$.  

In addition, note that the split-door method averages the estimate $\hat{\rho}$ obtained from each instance.
Not all instances may violate the connectedness assumption, therefore we introduce an additional sensitivity parameter $\kappa$ that denotes the fraction of invalid split-door instances. Bias in the final split-door estimate is then given by the following equation in the three sensitivity parameters:
\begin{equation}
    \operatorname{E}[\hat{\rho}] = \rho + \kappa c_1 c_2 \,.
\end{equation}

For expositional clarity, the above analysis assumed a linear structural model and demonstrated similarities with sensitivity of conditioning-based methods to unobserved common causes. However, in practice, the structural model may not be linear. In the recommendation example discussed in Section~\ref{sec:amazon}, we do not assume a linear model and instead use an aggregate ratio estimator. As shown in Figure~\ref{fig:sens-analysis}, simulations show that sensitivity of this estimator follows a similar bilinear dependence on $c_1$ and $c_2$.

\section{Characterizing error in the split-door estimate for a recommendation system}
\label{sec:error-bar}
In Section~\ref{sec:choose-tau}, the split-door causal estimate is defined as the mean of CTR estimates over all time periods and focal products with valid split-door instances. Here we characterize the error in this estimate. The key idea is that the error comes from two components: the first due to some erroneously identified split-door instances, and the second due to natural variance in estimating the mean. For a significance level $\alpha$ of the independence test, let $W$ be the number of obtained split-door instances and $N$ be the number of aggregated CTR estimates $\hat{\rho}_{i\tau}$ computed from these instances. 
Then the mean estimate can be written as:
\begin{equation}
\hat{\rho} = \frac{\sum_{i\tau}\hat{\rho}_{i\tau}}{N},
\end{equation}
where $i$ refers to a focal product and $\tau$ refers to a split-door time period. As in Appendix~\ref{sec:fdr}, let $\phi$ denote the expected fraction of erroneous split-door instances obtained. That is, for an expected number of $\phi W$ instances, the method may have erroneously concluded that the focal and recommended products are independent. Correspondingly, an expected $\phi W =\phi'N$ number of $\hat{\rho}_{i\tau}$ estimates will be invalid.\footnote{In general, the expected number of invalid $\hat{\rho}_{i\tau}$ estimates may be less than or equal to $\phi W$, since a focal product may have more than one recommended product that corresponds to an invalid split-door instance.} These invalid estimates can be expanded as:
\begin{equation}
\hat{\rho}_{i\tau}=\rho_{i\tau }^{causal}+\eta_{i\tau},
\end{equation}
where $\eta$ refers to the click-through rate due to correlated demand between  the focal and recommended products. Thus, the overall mean estimate can be written as:
\begin{align*}
    \hat{\rho} & =\frac{\sum_{i\tau \in A}\rho_{i\tau}^{causal} + \sum_{i\tau \in B}(\rho_{i\tau}^{causal}+\eta_{i\tau})}{N} \\
               & =\frac{\sum_{i\tau}\rho_{i\tau}^{causal}}{N} + \frac{\sum_{i\tau \in B}\eta_{i\tau}}{N},
\end{align*}
where $A$ and $B$ refer to $(i,\tau)$ pairs with valid and erroneous split-door estimates respectively ($|A|=(1-\phi')N, |B|=\phi' N$). 

Comparing this to the true $\rho^{causal}$, we obtain
\begin{equation}
\rho^{causal} - \hat{\rho}= (\rho^{causal}-\bar{\rho}^{causal}) - \frac{\sum_{i\tau \in B}\eta_{i\tau}}{N}\,.
\end{equation}
The first term of the RHS corresponds to error due to sampling variance, and the second term corresponds to error due to correlated demand ($\phi$). We estimate these terms below.
 
\paragraph{Error due to $\phi$}
Based on the argument for justifying the independence assumption in Section~\ref{sec:amz-build-model}, let us assume that the total effect of $U_Y$ on $Y_R$ is positive (without stipulating it for each individual instance). This means that the term due to correlated demand is positive, $\sum_{i\tau \in B}\eta_{i\tau} \geq 0$. Further, the maximum value of $\eta_{i\tau}$ is attained when all the observed click-throughs are due to correlated demand ($\eta_{i\tau}=\hat{\rho}_{i\tau}$). Under this assumption, 
$$0 \leq \frac{\sum_{i\tau \in B}\eta_{i\tau}}{N} \leq  \frac{\rho_{maxsum}}{N},$$
where $\rho_{maxsum}$ corresponds to the maximum sum of any subset of $\phi' N$ $\hat{\rho}_{i\tau}$ values. An approximate estimate can be derived using $\hat{\rho}$---the empirical mean over all $N$ values of $\rho_{i\tau}$---leading to $\rho_{maxsum} \approx \phi' N \hat{\rho}$.

\paragraph{Error due to natural variance}
We characterize this error by the 99\% confidence interval for the mean estimate, given by $2.58*\frac{\hat{\sigma}}{\sqrt{N}}$, where $\hat{\sigma}$ is the empirical standard deviation. 

Combining these two, the resultant interval for the split-door estimate is
\begin{equation} \label{eqn:bounds}
(\hat{\rho}-\frac{\rho_{maxsum}}{N}-2.58\frac{\hat{\sigma}}{\sqrt{N}}, \hat{\rho}+ 2.58\frac{\hat{\sigma}}{\sqrt{N}}) \,.
\end{equation}

The above interval demonstrates the bias-variance tradeoff in choosing a nominal significance level for the independence test and the corresponding $\phi$. At high nominal significance level $\alpha$, bias due to $\phi$ is expected to be low but variance of the estimate may be high due to low $N$. Conversely, at low values of $\alpha$, variance will be lower but $\phi$ is expected to be higher because we accept many more split-door instances.
